\documentclass[notitlepage,a4paper,aps,prd,onecolumn,superscriptaddress,nofootinbib,groupedaddress,longbibliography]{revtex4-1}
\usepackage{amsmath}
\usepackage{amsfonts}
\usepackage{amssymb}
\usepackage{mathtools}
\usepackage[utf8]{inputenc}
\usepackage[T1]{fontenc}
\usepackage[colorlinks=true]{hyperref}

\newcommand{\order}[2]{\overset{\mathclap{\scriptscriptstyle #2}}{#1}\vphantom{#1}}
\newcommand{\gauge}[2]{\prescript{\mathcal{#2}}{}{#1}}
\newcommand{\gaor}[3]{\prescript{\mathcal{#2}}{}{\vphantom{#1}}\overset{\mathclap{\scriptscriptstyle #3}}{#1}\vphantom{#1}}
\newcommand{\pert}[3]{\prescript{\mathcal{#2}}{}{#1}^{(#3)}}
\newcommand{\Lie}{\pounds}
\newcommand{\bdiamond}{\scriptscriptstyle\blacklozenge}

\begin{document}

\title{Gauge-invariant approach to the parametrized post-Newtonian formalism}

\author{Manuel Hohmann}
\email{manuel.hohmann@ut.ee}
\affiliation{Laboratory of Theoretical Physics, Institute of Physics, University of Tartu, W. Ostwaldi 1, 50411 Tartu, Estonia}

\begin{abstract}
We present an approach to the parametrized post-Newtonian (PPN) formalism which is based on gauge-invariant higher order perturbation theory. This approach divides the components of the metric perturbations into gauge-invariant quantities, which carry information about the physical system under consideration, and pure gauge quantities, which describe the choice of the coordinate system. This separation generally leads to a simplification of the PPN procedure, since only the gauge-invariant quantities appear in the field equations and must be determined by solving them. Another simplification arises from the fact that the gauge-invariant approach supersedes the necessity to first choose a gauge for solving the gravitational field equations and later transforming the obtained solution into the standard PPN gauge, as it is conventionally done in the PPN formalism, whose standard PPN gauge is determined only after the full solution is known. In addition to the usual metric formulation, we also present a tetrad formulation of the gauge-invariant PPN formalism. To illustrate their practical application, we demonstrate the calculation of the PPN parameters of a well-known scalar-tensor class of theories.
\end{abstract}

\maketitle


\section{Introduction}\label{sec:intro}
The open questions in cosmology and the tensions between general relativity and quantum theory have stipulated the study of a plethora of modified theories of gravity. Besides addressing these open problems, any such theory must of course also conform with the numerous tests of gravity performed in laboratory experiments, in the solar system as well as using observations of the orbital motion of extrasolar objects. An indispensable tool for testing the viability of gravity theories using this set of high-precision data is the parametrized post-Newtonian (PPN) formalism~\cite{Nordtvedt:1968qs,Thorne:1970wv,Will:1971zzb,Will:1971wt,Will:1993ns,Will:2014kxa,Will:2018bme}. It allows to characterize any given theory of gravity which satisfies a number of assumptions, such as the existence of a metric governing the motion of test bodies, by a set of (usually constant) parameters, which can be derived from a perturbative solution of the field equations in a weak-field approximation. These parameters can then be compared to observations, e.g., in the solar system.

A basic assumption of the PPN formalism is the existence of a distinguished coordinate system, conventionally identified with the universe rest frame, in which the gravitational field is given by a perturbation of a fixed background metric, usually the flat Minkowski metric. The PPN formalism then prescribes to expand the field equations around this background solution up to quadratic order in the perturbations, and to solve for the perturbations order by order, where at each level of the perturbation theory the equations to be solved are linear in the unknowns. While this procedure is straightforward in principle, it may pose practical difficulties if the field equations involve non-trivial couplings between the metric and possibly other tensor fields, intertwining their components in a way which makes them difficult to separate. Another issue arises from the remaining diffeomorphism invariance, which stems from the fact that the PPN coordinate system is a priori defined only up to coordinate transformations which are of the order of the metric perturbations, and is fully determined only after solving the field equations. This results in both the freedom and the necessity to choose the coordinates during the process of solving the field equations, by supplementing them with an arbitrary choice of gauge conditions on the metric perturbations, and possibly adapting this choice once the solution is obtained, and there is in general no canonical way to make this choice.

While the assumption of a distinguished coordinate system describing the universe rest frame may be reasonable from an experimental point of view, it appears at least unnatural from a point of view which attributes the gravitational interaction to the geometry of spacetime. In this picture, an important role is given to diffeomorphism invariance, and hence the independence of the choice of coordinate systems. One may therefore ask whether it is possible to interpret the PPN formalism and its use of perturbation theory in alignment with this geometric point of view, with the fixed coordinate system merely being an artifact of its conventional formulation, while at the same time resolving the issues of non-canonical gauge choices and possibly cumbersome equations to be solved.

Taking a look aside towards cosmology, in which perturbation theory likewise plays an important role to link theory and experiment, one finds the gauge-invariant theory of linear perturbations~\cite{Bardeen:1980kt,Kodama:1985bj,Mukhanov:1990me,Malik:2008im} as a common tool. In order to devise a similar gauge-invariant approach to the PPN formalism, one must go beyond the linear approximation and consider higher order gauge-invariant perturbations. For this purpose we make use of the theory of nonlinear gauge transformations~\cite{Bruni:1996im,Bruni:1999et}, from which a higher order Taylor expansion of tensor fields was obtained~\cite{Sonego:1997np}. This work has been extended to perturbations in more than one variable~\cite{Nakamura:2003wk,Nakamura:2004wr}, and likewise been applied in the context of cosmology~\cite{Nakamura:2004rm,Nakamura:2006rk}, thus providing a gauge-invariant formulation to numerous preceding studies of higher order cosmological perturbations~\cite{Tomita:PTP.37.831,Tomita:PTP.45.1747,Tomita:PTP.47.416}.

The aim of this article is to make use of the gauge-invariant higher order perturbation theory mentioned above and to apply it to the PPN formalism, in order to address the aforementioned potential difficulties and to provide a fundamentally geometric interpretation. This requires addressing a few peculiarities of the PPN formalism, such as assigning different perturbation orders to space and time derivatives, and to explicitly derive the gauge transformations and gauge-invariant description of various quantities appearing in the PPN formalism, such as the matter energy-momentum and the fundamental fields mediating the gravitational interaction. To allow for an easy and straightforward application of our formulation of the PPN formalism in practice, we provide explicit formulas whenever it appears useful.

The outline of this article is as follows. In section~\ref{sec:gipert} we provide a brief review of higher order gauge-invariant perturbation theory, with particular focus on our intention to apply it to the PPN formalism. The gauge-invariant PPN formalism is then developed in section~\ref{sec:gippn}, using the conventional metric formulation. An alternative approach based on tetrads, which is more suitable for certain classes of gravity theories, is presented in section~\ref{sec:tetrad}. To illustrate the use of our formalism, we apply it to a simple, yet non-trivial example in section~\ref{sec:example}. We end with a conclusion in section~\ref{sec:conclusion}. Throughout this article we use Greek letters \(\mu, \nu = 0, \ldots, 3\) to denote spacetime indices, while Latin letters \(i, j = 1, \ldots, 3\) denote spatial indices.

\section{Gauge-invariant perturbation theory}\label{sec:gipert}
We start with a brief review of higher order gauge-invariant perturbation theory~\cite{Bruni:1996im,Bruni:1999et,Sonego:1997np,Nakamura:2003wk,Nakamura:2004wr,Nakamura:2004rm,Nakamura:2006rk}, where we focus on the aspects which will be important for our intended application to the PPN formalism. First, we give a fully geometric definition of the notion of gauge we use in section~\ref{ssec:gaugedef}. Making use of this notion, we proceed to the definition of perturbations in section~\ref{ssec:pertdef}. Gauge transformations in this picture are discussed in section~\ref{ssec:trafodef}. Finally, in section~\ref{ssec:ginvdef}, we define the notion of gauge-invariant quantities. Although we explicitly mention only the metric in this section, the same procedure applies to any other tensor fields present in a particular gravitational theory under consideration.

\subsection{Definition of gauge}\label{ssec:gaugedef}
We start our discussion of gauge invariance with a general remark on the use of gauging in the literature. In the context of diffeomorphism invariance, the term \emph{gauge} is often used synonymously to denote a choice of coordinates, hence a chart of the spacetime manifold, which is then used to express the components of tensor fields. Following this interpretation, gauge transformations are represented by changes of coordinates. This corresponds to what is known as passive interpretation of a diffeomorphism: points on the manifold and tensor fields at these points are the same, but the labels given to these points and the tensor components with respect to this labeling change. For our purposes, however, it will turn out to be more convenient to resort to the active interpretation of diffeomorphisms: a fixed coordinate system is chosen, points are mapped to a different position and tensor fields are moved and changed along with them. We will make this notion mathematically precise below, following the definitions given in~\cite{Bruni:1999et,Nakamura:2004wr}.

Let \(M_0\) be a manifold equipped with a metric \(g^{(0)}\). We call \((M_0, g^{(0)})\) the background spacetime, and \(g^{(0)}\) the background metric. We choose this background spacetime to be some ``standard'' spacetime equipped with a fixed choice of coordinates. Common examples are Minkowski space with Cartesian coordinates or other maximally symmetric spacetimes. The background spacetime will serve as a reference, to which we can compare a second, different manifold \(M\) equipped with a metric \(g\), which we call the physical spacetime and physical metric, respectively. However, we cannot immediately compare the metrics \(g\) and \(g^{(0)}\) for two reasons:
\begin{enumerate}
\item
The metrics \(g\) and \(g^{(0)}\) are defined on two different manifolds. There is no notion of a ``background metric'' on the physical spacetime, since the background metric is defined only on the reference spacetime.
\item
There is no canonical identification between points of the physical and reference spacetimes. In other words, there is no canonical choice of coordinates on the physical spacetime.
\end{enumerate}
In order to compare the two metrics, we must therefore choose a diffeomorphism \(\mathcal{X}: M_0 \to M\). This diffeomorphism will do two things:
\begin{enumerate}
\item
It allows us to identify points on the manifolds \(M_0\) and \(M\). Hence, it will equip \(M\) with a distinguished choice of coordinates, obtained from the coordinates on \(M_0\).
\item
It defines the pullback \(\mathcal{X}^*g\) of \(g\) to \(M_0\), which we will also write as \(\gauge{g}{X}\), and which we can compare to the background metric \(g^{(0)}\).
\end{enumerate}
The diffeomorphism \(\mathcal{X}\) is what we will call a \emph{gauge}~\cite{Bruni:1999et,Nakamura:2004wr}. Note that there is no canonical choice for such a diffeomorphism.

\subsection{Perturbations}\label{ssec:pertdef}
In perturbation theory we assume that the physical metric depends on a parameter \(\epsilon\), commonly called the perturbation parameter, and so we will denote it by \(g_{\epsilon}\). We also assume that for each value of \(\epsilon\), the metric is defined on a different physical spacetime \(M_{\epsilon}\). In order to compare the physical and background metrics, we therefore need a family of diffeomorphisms \(\mathcal{X}_{\epsilon}: M_0 \to M_{\epsilon}\). Further, we assume that the ``unperturbed'' metric \(g_0\) agrees with the background metric \(g^{(0)}\) on \(M_0\). Hence, for consistency we assume \(\mathcal{X}_0 = \mathrm{id}_{M_0}: M_0 \to M_0\) is the identity map.\footnote{Note that in contrast to the treatment in~\cite{Nakamura:2004wr,Nakamura:2004rm,Nakamura:2006rk} we will not regard the manifolds \(M_{\epsilon}\) as leaves of a foliation of a manifold \(N \cong M \times \mathbb{R}\), as it will not be necessary for the construction we present here.}

A key idea of perturbation theory is the assumption that the physical metric \(g_{\epsilon}\) can be approximated by a series expansion in the perturbation parameter \(\epsilon\), whose zeroth order is the background metric \(g^{(0)}\). Following our discussion above, we see that we cannot perform such a series expansion directly, since \(g_{\epsilon}\) and \(g^{(0)}\) are defined on different manifolds. We can, however, express the pullback \(\gauge{g}{X}_{\epsilon} = \mathcal{X}_{\epsilon}^*g_{\epsilon}\), which is defined on \(M_0\), as a series expansion of the form
\begin{equation}\label{eq:perturb}
\gauge{g}{X}_{\epsilon} = \sum_{k = 0}^{\infty}\frac{\epsilon^k}{k!}\left.\frac{\partial^k\gauge{g}{X}_{\epsilon}}{\partial\epsilon^k}\right|_{\epsilon = 0} = \sum_{k = 0}^{\infty}\frac{\epsilon^k}{k!}\pert{g}{X}{k}\,.
\end{equation}
Clearly, \(\pert{g}{X}{0} = g^{(0)}\) is the background metric, and hence independent of the choice of the gauge \(\mathcal{X}_{\epsilon}\). The other series coefficients \(\pert{g}{X}{k}\) for \(k \geq 1\), however, will depend on the choice of the gauge.

\subsection{Gauge transformations}\label{ssec:trafodef}
We now consider two different gauges \(\mathcal{X}_{\epsilon}, \mathcal{Y}_{\epsilon}: M_0 \to M_{\epsilon}\). This allows us to construct a family \(\Phi_{\epsilon}: M_0 \to M_0\) of diffeomorphisms given by \(\Phi_{\epsilon} = \mathcal{X}_{\epsilon}^{-1} \circ \mathcal{Y}_{\epsilon}\). The metrics \(\gauge{g}{X}_{\epsilon} = \mathcal{X}_{\epsilon}^*g, \gauge{g}{Y}_{\epsilon} = \mathcal{Y}_{\epsilon}^*g\) in the different gauges, which are now both defined on the background spacetime \(M_0\), are related by
\begin{equation}
\gauge{g}{Y}_{\epsilon} = \Phi_{\epsilon}^*\gauge{g}{X}_{\epsilon}\,.
\end{equation}
Note that \(\Phi\) is only a one-parameter \emph{family} of diffeomorphisms, but in general not a one-parameter \emph{group}; one has \(\Phi_{\epsilon + \epsilon'} \neq \Phi_{\epsilon} \circ \Phi_{\epsilon'}\) and \(\Phi_{-\epsilon} \neq \Phi_{\epsilon}^{-1}\) in general. However, one can show that for any one-parameter family \(\Phi\) of diffeomorphisms there exists an (in general infinite) series of one-parameter groups \(\phi^{(k)}\) of diffeomorphisms such that~\cite{Bruni:1996im}
\begin{equation}
\Phi_{\epsilon} = \cdots \phi^{(k)}_{\epsilon^k/k!} \circ \cdots \circ \phi^{(2)}_{\epsilon^2/2} \circ \phi^{(1)}_{\epsilon}\,.
\end{equation}
Since each \(\phi^{(k)}\) is a one-parameter group of diffeomorphisms, it is generated by a vector field, which we will denote by \(\xi_{(k)}\). It turns out that the metrics in the two different gauges are related by the series expansion
\begin{equation}
\gauge{g}{Y}_{\epsilon} = \sum_{l_1 = 0}^{\infty}\cdots\sum_{l_j = 0}^{\infty}\cdots\frac{\epsilon^{l_1 + \cdots + jl_j + \cdots}}{(1!)^{l_1} \cdots (j!)^{l_j} \cdots l_1! \cdots l_j! \cdots}\Lie_{\xi_{(1)}}^{l_1} \cdots \Lie_{\xi_{(j)}}^{l_j} \cdots \gauge{g}{X}_{\epsilon}\,.
\end{equation}
The coefficients of the Taylor expansion are thus related by
\begin{equation}
\pert{g}{Y}{k} = \left.\frac{\partial^k\gauge{g}{Y}_{\epsilon}}{\partial\epsilon^k}\right|_{\epsilon = 0} = \sum_{0 \leq l_1 + 2l_2 + \ldots \leq k}\frac{k!}{(k - l_1 - 2l_2 - \ldots)!(1!)^{l_1}(2!)^{l_2} \cdots l_1!l_2! \cdots}\Lie_{\xi_{(1)}}^{l_1} \cdots \Lie_{\xi_{(j)}}^{l_j} \cdots \pert{g}{X}{k - l_1 - 2l_2 - \ldots}\,.
\end{equation}
Writing out the lowest four orders of this formula we find
\begin{subequations}\label{eq:pertgtrans}
\begin{align}
\pert{g}{Y}{0} &= \pert{g}{X}{0} = g_0\,,\\
\pert{g}{Y}{1} &= \pert{g}{X}{1} + \Lie_{\xi_{(1)}}\pert{g}{X}{0}\,,\\
\pert{g}{Y}{2} &= \pert{g}{X}{2} + 2\Lie_{\xi_{(1)}}\pert{g}{X}{1} + \Lie_{\xi_{(2)}}\pert{g}{X}{0} + \Lie_{\xi_{(1)}}^2\pert{g}{X}{0}\,,\\
\pert{g}{Y}{3} &= \pert{g}{X}{3} + 3\Lie_{\xi_{(1)}}\pert{g}{X}{2} + 3\Lie_{\xi_{(2)}}\pert{g}{X}{1} + 3\Lie_{\xi_{(1)}}^2\pert{g}{X}{1} + \Lie_{\xi_{(3)}}\pert{g}{X}{0} + 3\Lie_{\xi_{(1)}}\Lie_{\xi_{(2)}}\pert{g}{X}{0} + \Lie_{\xi_{(1)}}^3\pert{g}{X}{0}\,,\\
\pert{g}{Y}{4} &= \pert{g}{X}{4} + 4\Lie_{\xi_{(1)}}\pert{g}{X}{3} + 6\Lie_{\xi_{(2)}}\pert{g}{X}{2} + 6\Lie_{\xi_{(1)}}^2\pert{g}{X}{2} + 4\Lie_{\xi_{(3)}}\pert{g}{X}{1} + 12\Lie_{\xi_{(1)}}\Lie_{\xi_{(2)}}\pert{g}{X}{1}\nonumber\\
&\phantom{=}+ 4\Lie_{\xi_{(1)}}^3\pert{g}{X}{1} + \Lie_{\xi_{(4)}}\pert{g}{X}{0} + 3\Lie_{\xi_{(2)}}^2\pert{g}{X}{0} + 4\Lie_{\xi_{(1)}}\Lie_{\xi_{(3)}}\pert{g}{X}{0} + 6\Lie_{\xi_{(1)}}^2\Lie_{\xi_{(2)}}\pert{g}{X}{0} + \Lie_{\xi_{(1)}}^4\pert{g}{X}{0}\,.
\end{align}
\end{subequations}
Observe that for each term on the right hand side the perturbation order, given by the sum of the perturbation orders of \(\pert{g}{X}{k}\) and the vector fields \(\xi_{(k)}\) in the Lie derivatives, agrees with the perturbation order of the left hand side.

\subsection{Gauge-invariant quantities}\label{ssec:ginvdef}
The main idea of gauge-invariant perturbation theory is to divide the variables describing the metric \(\gauge{g}{X}_{\epsilon}\) into gauge independent variables \(\mathbf{g}_{\epsilon}\), which describe properties of the physical metric \(g_{\epsilon}\) and hence observable quantities, and gauge dependent variables, which describe properties of the gauge only~\cite{Nakamura:2004wr,Nakamura:2004rm,Nakamura:2006rk}. One possibility to achieve this separation and to introduce gauge invariant quantities is to choose a distinguished gauge \(\mathcal{S}_{\epsilon}\). This gauge can be obtained, for example, by imposing gauge conditions on the metric, such as the standard post-Newtonian or harmonic gauges. If these uniquely fix the gauge \(\mathcal{S}_{\epsilon}\), we may use it to define the gauge invariant metric
\begin{equation}
\mathbf{g}_{\epsilon} = \mathcal{S}_{\epsilon}^*g_{\epsilon}\,.
\end{equation}
Given any other gauge \(\mathcal{X}_{\epsilon}\), we can write the metric in this gauge in the form \(\gauge{g}{X}_{\epsilon} = \mathcal{X}_{\epsilon}^*(\mathcal{S}_{\epsilon}^{-1})^*\mathbf{g}_{\epsilon}\), i.e., by applying a gauge transformation. We have thus achieved a split of \(\gauge{g}{X}_{\epsilon}\) into a gauge dependent part, namely the diffeomorphism \(\mathcal{S}_{\epsilon}^{-1} \circ \mathcal{X}_{\epsilon}\), describing the gauge, and a gauge-invariant part \(\mathbf{g}_{\epsilon}\), describing the physical metric. Note that this also implies a split of the number of free components of \(\gauge{g}{X}_{\epsilon}\): the gauge invariant metric \(\mathbf{g}_{\epsilon}\) has fewer free components, since some components are fixed by the gauge conditions corresponding to the choice of the distinguished gauge \(\mathcal{S}_{\epsilon}\). These missing components are exactly found in the gauge transformations, if a different gauge \(\mathcal{X}_{\epsilon}\) is chosen. We will illustrate this fact when de define the gauge invariant PPN metric in section~\ref{ssec:ginvmetric}.

As discussed above, for any gauge transformation there exist vector fields, which we will now denote by \(X_{(k)}\), such that the metric can be written as a series expansion
\begin{equation}\label{eq:ginvexpan}
\gauge{g}{X}_{\epsilon} = \sum_{l_1 = 0}^{\infty}\cdots\sum_{l_j = 0}^{\infty}\cdots\frac{\epsilon^{l_1 + \cdots + jl_j + \cdots}}{(1!)^{l_1} \cdots (j!)^{l_j} \cdots l_1! \cdots l_j! \cdots}\Lie_{X_{(1)}}^{l_1} \cdots \Lie_{X_{(j)}}^{l_j} \cdots \mathbf{g}_{\epsilon}\,.
\end{equation}
Also the series expansion coefficients are related by
\begin{equation}\label{eq:ginvcoeff}
\pert{g}{X}{k} = \sum_{0 \leq l_1 + 2l_2 + \ldots \leq k}\frac{k!}{(k - l_1 - 2l_2 - \ldots)!(1!)^{l_1}(2!)^{l_2} \cdots l_1!l_2! \cdots}\Lie_{X_{(1)}}^{l_1} \cdots \Lie_{X_{(j)}}^{l_j} \cdots \mathbf{g}^{(k - l_1 - 2l_2 - \ldots)}\,.
\end{equation}
The Taylor coefficients \(\pert{g}{X}{k}\) thus also split into a gauge dependent part \(X_{(k)}\) and a gauge-invariant part \(\mathbf{g}^{(k)}\). Given any other gauge \(\mathcal{Y}_{\epsilon}\), the same formula holds for a different family \(Y_{(k)}\) of vector fields, but with the same gauge invariant part \(\mathbf{g}^{(k)}\). The gauge defining vector fields \(X_{(k)}\) and \(Y_{(k)}\) are thus the only components of this split which change under a gauge transformation \(\Phi_{\epsilon} = \mathcal{X}_{\epsilon}^{-1} \circ \mathcal{Y}_{\epsilon}\). Writing the generating vector fields of \(\Phi_{\epsilon}\) as \(\xi_{(k)}\), one finds that the transformation is given by
\begin{subequations}
\begin{align}
Y_{(1)} &= X_{(1)} + \xi_{(1)}\,,\\
Y_{(2)} &= X_{(2)} + \xi_{(2)} + [\xi_{(1)}, X_{(1)}]\,,\\
Y_{(3)} &= X_{(3)} + \xi_{(3)} + 3[\xi_{(2)}, X_{(1)}] - [\xi_{(1)}, [\xi_{(1)}, X_{(1)}]] + 2[[\xi_{(1)}, X_{(1)}], X_{(1)}]\,,\\
Y_{(4)} &= X_{(4)} + \xi_{(4)} + 3[\xi_{(2)}, X_{(2)}] + 4[\xi_{(3)}, X_{(1)}] + 6[[\xi_{(2)}, X_{(1)}], X_{(1)}] + 3[[\xi_{(1)}, X_{(1)}], X_{(2)}] - 3[\xi_{(2)}, [\xi_{(1)}, X_{(1)}]]\nonumber\\
&\phantom{=}+ [\xi_{(1)}, [\xi_{(1)}, [\xi_{(1)}, X_{(1)}]]] + 3[[[\xi_{(1)}, X_{(1)}], X_{(1)}], X_{(1)}] - 3[[\xi_{(1)}, [\xi_{(1)}, X_{(1)}]], X_{(1)}]
\end{align}
\end{subequations}
and similarly for higher orders. The particular form of the gauge-invariant perturbations, of course, depends on the choice of the standard gauge \(\mathcal{S}_{\epsilon}\), which must be adapted to the particular problem under consideration. We will show one possibility how to make this gauge choice when we apply the gauge invariant perturbation theory to the PPN formalism in the following section.

\section{Gauge-invariant PPN formalism}\label{sec:gippn}
With the necessary mathematical background at hand, we may now develop a gauge-invariant approach to the PPN formalism, where the notion of gauge is to be understood as discussed in the preceding section. For this purpose, we first review the notion of PPN perturbation orders and its associated peculiarities in section~\ref{ssec:ppnpert}. We then discuss post-Newtonian gauge transformations in section~\ref{ssec:gtrans}. We use them to define the gauge-invariant metric perturbations in section~\ref{ssec:ginvmetric}. In section~\ref{ssec:enmomtens}, we apply the gauge-invariant description to the energy-momentum tensor of a perfect fluid, which will act as the matter source. This matter source is further described in terms of the so-called PPN potentials in section~\ref{ssec:ppnpot}. Finally, in section~\ref{ssec:standardppn} we consider at the standard PPN form of the metric, and decompose it into its gauge-invariant and gauge dependent parts. This step will finally allow us to determine the PPN parameters from the gauge-invariant metric components, which can then be compared to observations. Our treatment uses the definitions and notation used in~\cite[Sec. 4]{Will:1993ns}; a slightly modified treatment is presented in~\cite[Sec. 4]{Will:2018bme}.

\subsection{Post-Newtonian perturbation orders}\label{ssec:ppnpert}
In order to derive a gauge-invariant approach to the PPN formalisms, it is important to notice a few peculiarities about its use of perturbation theory, compared to the standard theory we have discussed in the previous section. Most of these peculiarities arise from the assumption that the matter which acts as the source of the gravitational field is given by a perfect fluid, whose velocity in a particular, fixed coordinate system \((x^{\mu})\), usually identified with the ``universe rest frame'', is small, measured in units of the speed of light, and that this velocity acts as the perturbation parameter. On the physical spacetime \(M\) this fluid is described by a rest energy density \(\rho\), specific internal energy \(\Pi\), pressure \(p\) and four-velocity \(u^{\mu}\), so that its energy-momentum tensor is given by
\begin{equation}\label{eq:tmunu}
T^{\mu\nu} = (\rho + \rho\Pi + p)u^{\mu}u^{\nu} + pg^{\mu\nu}\,.
\end{equation}
The four-velocity \(u^{\mu}\) is normalized by the metric \(g_{\mu\nu}\), so that \(u^{\mu}u^{\nu}g_{\mu\nu} = -1\). Following our treatment in the previous section, coordinates are introduced as a diffeomorphism \(\mathcal{X}: M_0 \to M\) from a given reference spacetime \(M_0\) to the physical spacetime, and replacing all tensorial quantities mentioned above by their pullbacks \(\gauge{\bullet}{X} = \mathcal{X}^*\bullet\) along \(\mathcal{X}\). We then assume that the velocity \(\gauge{v}{X}^i = \gauge{u}{X}^i/\gauge{u}{X}^0\) of the source matter in these coordinates is small, \(|\gauge{\vec{v}}{X}| = \epsilon \ll 1\), so that it may serve as a perturbation parameter. Hence, all tensor perturbations will be measured in terms of velocity orders \(\mathcal{O}(n) \propto |\gauge{\vec{v}}{X}|^n\).

One of the aforementioned peculiarities is the fact that time derivatives are weighted with an additional velocity order, \(\partial_0 \sim \mathcal{O}(1)\). This can be implemented in the standard perturbation theory by choosing the coordinates to be \((x^0 = ct, x^i)\) and taking \(c^{-1}\) as the perturbation parameter. Then one naturally obtains
\begin{equation}
\frac{\partial}{\partial x^0} = \frac{1}{c}\frac{\partial}{\partial t} \quad \Rightarrow \quad \frac{\partial}{\partial x^0}\mathcal{O}(n) \sim \frac{\partial}{\partial t}\mathcal{O}(n + 1)\,,
\end{equation}
while spatial derivatives retain the perturbation order. Thus, one may keep track of perturbation orders in this way by counting powers of \(c\). However, this is rather tedious, and so we will omit this step here, since the relevant perturbation orders have been thoroughly worked out for the PPN formalism~\cite{Will:1993ns}. Further, it is conventional to write the perturbation expansion in the form
\begin{equation}\label{eq:metppnexp}
\gauge{g}{X} = \sum_{k = 0}^{\infty}\gaor{g}{X}{k}\,.
\end{equation}
Comparing this with our previous definition~\eqref{eq:perturb} we see that the factor \(\epsilon^k/k!\) has been absorbed into the perturbation \(\gaor{g}{X}{k} \sim \mathcal{O}(k)\). We also suppressed the perturbation parameter \(\epsilon\) in this notation. Here the zeroth order is given by the background metric, which we assume to be a flat Minkowski background, \(\gaor{g}{X}{0}_{\mu\nu} = \eta_{\mu\nu} = \mathrm{diag}(-1, 1, 1, 1)\). Note that also more general choices, such as a Friedmann-Lemaitre-Robertson-Walker metric, are possible, but negligible for effects on solar system scales~\cite[Sec. 4.1.3]{Will:2018bme}.

In order to determine which terms in the expansion~\eqref{eq:metppnexp} are relevant, one considers the action
\begin{equation}\label{eq:partaction}
S[\gamma] = -m\int\sqrt{-\gauge{g}{X}_{\mu\nu}\gauge{\dot{\gamma}}{X}^{\mu}\gauge{\dot{\gamma}}{X}^{\nu}}dt = -m\int\sqrt{-\gauge{g}{X}_{00} - 2\gauge{g}{X}_{0i}\gauge{\dot{\gamma}}{X}^i - \gauge{g}{X}_{ij}\gauge{\dot{\gamma}}{X}^i\gauge{\dot{\gamma}}{X}^j}dt
\end{equation}
of a test particle of mass \(m\) moving along a trajectory \(\gauge{\gamma}{X}^{\mu} = (t,\vec{x})\), which we chose to parametrize by coordinate time \(t\) in the gauge \(\mathcal{X}\). One can then distinguish two different cases:
\begin{enumerate}
\item
In the first case, one assumes that the velocity of the test particle is of the same order \(|\gauge{\dot{\vec{\gamma}}}{X}| \sim \mathcal{O}(1)\) as that of the source matter. Expanding the action~\eqref{eq:partaction} into velocity orders one finds that the second velocity order corresponds to the Newtonian limit, with \(\gaor{g}{X}{2}_{00} = 2\gauge{U}{X}\) given by the Newtonian potential, while the post-Newtonian limit requires to consider terms up to the fourth velocity order, which contain the metric components \(\gaor{g}{X}{2}_{ij}, \gaor{g}{X}{3}_{0i}, \gaor{g}{X}{4}_{00}\).

\item
The second case is given by assuming that the velocity of the test particle is of the order \(1\), which is the case if one studies the propagation of light. Here one finds that the first post-Newtonian correction appears already at the second velocity order, while the fourth velocity order yields a second post-Newtonian correction~\cite{Richter:1982zz,Richter:1982zza}. In this case one expands the metric up to the orders \(\gaor{g}{X}{4}_{ij}, \gaor{g}{X}{4}_{0i}, \gaor{g}{X}{4}_{00}\).
\end{enumerate}
In addition to these considerations, which determine the maximal velocity orders we consider, the appearing metric components are further restricted by two more physical considerations:
\begin{enumerate}
\item
As we will discuss in section~\ref{ssec:enmomtens}, the lowest order components of the energy-momentum tensor, which is the source of the gravitational field equations, are of the second velocity order. Hence, no metric perturbations at the first velocity order appear, \(\gaor{g}{X}{1} = 0\).

\item
The terms involving perturbations \(\gaor{g}{X}{2}_{0i}, \gaor{g}{X}{3}_{00}, \gaor{g}{X}{3}_{ij}, \gaor{g}{X}{4}_{0i}\) in the action~\eqref{eq:partaction} contain an odd number of velocity factors, so that they are antisymmetric under time reversal, and thus correspond to dissipative processes. These are prohibited by energy-momentum conservation: conservation of rest mass prohibits terms of the first velocity order, while the third velocity order is prohibited by Newtonian energy conservation~\cite{Will:1993ns}.
\end{enumerate}
In summary, we will therefore consider in the following only the components
\begin{equation}\label{eq:metcompgauge}
\gaor{g}{X}{2}_{00}\,, \quad
\gaor{g}{X}{2}_{ij}\,, \quad
\gaor{g}{X}{3}_{0i}\,, \quad
\gaor{g}{X}{4}_{00}\,, \quad
\gaor{g}{X}{4}_{ij}\,.
\end{equation}
The final component \(\gaor{g}{X}{4}_{ij}\) is usually not considered in the PPN formalism, since its contribution to the equations of motion of slow-moving test matter is subleading. However, it appears in general in the fourth order field equations of gravity theories and couples to the relevant component \(\gaor{g}{X}{4}_{00}\), and it may be used to calculate higher-order contributions to light deflection~\cite{Richter:1982zz,Richter:1982zza}, so that we will keep it here for completeness.

\subsection{Gauge transformations}\label{ssec:gtrans}
The conditions given above do not determine the coordinate system uniquely, but allow for a particular set of gauge transformations. These are restricted by the conditions that they retain the post-Newtonian character of the metric. In analogy to the metric components, we introduce the notation \(\order{\xi}{k} = \xi_{(k)}\epsilon^k/k!\) for the generating vector fields of the gauge transformation. One then finds that the only allowed and relevant components we have to consider are given by
\begin{equation}\label{eq:gatravec}
\order{\xi}{2}_i\,, \quad
\order{\xi}{3}_0\,, \quad
\order{\xi}{4}_i\,,
\end{equation}
where here and in the remainder of this article we use the Minkowski metric \(\eta\) to raise and lower indices of tensor fields on the background spacetime \(M_0\). Applying this gauge transformation to the metric components~\eqref{eq:metcompgauge} by using the general formula~\eqref{eq:pertgtrans} we find that in a different gauge \(\mathcal{Y}\) they take the form
\begin{subequations}\label{eq:ppnmetgatra}
\begin{align}
\gaor{g}{Y}{2}_{00} &= \gaor{g}{X}{2}_{00}\,,\\
\gaor{g}{Y}{2}_{ij} &= \gaor{g}{X}{2}_{ij} + 2\partial_{(i}\order{\xi}{2}_{j)}\,,\\
\gaor{g}{Y}{3}_{0i} &= \gaor{g}{X}{3}_{0i} + \partial_i\order{\xi}{3}_0 + \partial_0\order{\xi}{2}_i\,,\\
\gaor{g}{Y}{4}_{00} &= \gaor{g}{X}{4}_{00} + 2\partial_0\order{\xi}{3}_0 + \order{\xi}{2}_i\partial_i\gaor{g}{X}{2}_{00}\,,\\
\gaor{g}{Y}{4}_{ij} &= \gaor{g}{X}{4}_{ij} + 2\partial_{(i}\order{\xi}{4}_{j)} + 2\gaor{g}{X}{2}_{k(i}\partial_{j)}\order{\xi}{2}_k + \order{\xi}{2}_k\partial_k\gaor{g}{X}{2}_{ij} + \partial_{(i}(\order{\xi}{2}_{|k}\partial_{k|}\order{\xi}{2}_{j)}) + \partial_i\order{\xi}{2}_k\partial_j\order{\xi}{2}_k\,.
\end{align}
\end{subequations}
By a suitable choice of \(\xi\) it is possible to eliminate certain components from the metric. We will do so below, when we define the gauge-invariant metric components.

\subsection{Gauge-invariant metric}\label{ssec:ginvmetric}
Examining the gauge transformation~\eqref{eq:ppnmetgatra}, we see that by a suitable choice of \(\xi_i\) it is possible to eliminate certain components of \(g_{ij}\), such that only a diagonal (pure trace) and a trace-free, divergence-free part remain. Similarly, we may choose \(\xi_0\) such that any divergence is eliminated from \(g_{0i}\), and retain only a divergence-free part. These conditions uniquely fix a gauge, so that the remaining components, which are also uniquely determined independent of the gauge in which the metric was originally given, become gauge-invariant quantities. These can be parametrized in the form
\begin{equation}\label{eq:gimetric}
\mathbf{g}_{00} = \mathbf{g}^{\star}\,, \quad
\mathbf{g}_{0i} = \mathbf{g}^{\diamond}_i\,, \quad
\mathbf{g}_{ij} = \mathbf{g}^{\bullet}\delta_{ij} + \mathbf{g}^{\dagger}_{ij}
\end{equation}
by two scalars \(\mathbf{g}^{\star}, \mathbf{g}^{\bullet}\), a divergence-free vector \(\mathbf{g}^{\diamond}_i\) and a symmetric, trace-free, divergence-free tensor \(\mathbf{g}^{\dagger}_{ij}\). Note that we have used filled symbols to denote scalars, empty symbols for vectors and symbols without interior for tensors. These components satisfy the restrictions
\begin{equation}
\partial^i\mathbf{g}^{\diamond}_i = 0\,, \quad
\partial^i\mathbf{g}^{\dagger}_{ij} = 0\,, \quad
\mathbf{g}^{\dagger}_{[ij]} = 0\,, \quad
\mathbf{g}^{\dagger}_{ii} = 0\,.
\end{equation}
Making use of the series expansion~\eqref{eq:ginvcoeff}, we can expand the metric components in any arbitrary gauge \(\mathcal{X}\) as
\begin{subequations}\label{eq:trametric}
\begin{align}
\gaor{g}{X}{2}_{00} &= \order{\mathbf{g}}{2}^{\star}\,,\label{eq:trametric200}\\
\gaor{g}{X}{2}_{ij} &= \order{\mathbf{g}}{2}^{\bullet}\delta_{ij} + \order{\mathbf{g}}{2}^{\dagger}_{ij} + 2\partial_i\partial_j\order{X}{2}^{\bdiamond} + 2\partial_{(i}\order{X}{2}^{\diamond}_{j)}\,,\label{eq:trametric2ij}\\
\gaor{g}{X}{3}_{0i} &= \order{\mathbf{g}}{3}^{\diamond}_i + \partial_i\order{X}{3}^{\star} + \partial_0\partial_i\order{X}{2}^{\bdiamond} + \partial_0\order{X}{2}^{\diamond}_i\,,\label{eq:trametric30i}\\
\gaor{g}{X}{4}_{00} &= \order{\mathbf{g}}{4}^{\star} + 2\partial_0\order{X}{3}^{\star} + (\partial_i\order{X}{2}^{\bdiamond} + \order{X}{2}^{\diamond}_i)\partial_i\order{\mathbf{g}}{2}^{\star}\,,\label{eq:trametric400}\\
\gaor{g}{X}{4}_{ij} &= \order{\mathbf{g}}{4}^{\bullet}\delta_{ij} + \order{\mathbf{g}}{4}^{\dagger}_{ij} + 2\partial_i\partial_j\order{X}{4}^{\bdiamond} + 2\partial_{(i}\order{X}{4}^{\diamond}_{j)} + [\order{\mathbf{g}}{2}^{\bullet}\delta_{ik} + \order{\mathbf{g}}{2}^{\dagger}_{ik}]\partial_j(\partial_k\order{X}{2}^{\bdiamond} + \order{X}{2}^{\diamond}_k) + [\order{\mathbf{g}}{2}^{\bullet}\delta_{jk} + \order{\mathbf{g}}{2}^{\dagger}_{jk}]\partial_i(\partial_k\order{X}{2}^{\bdiamond} + \order{X}{2}^{\diamond}_k)\nonumber\\
&\phantom{=}+ (\partial_k\order{X}{2}^{\bdiamond} + \order{X}{2}^{\diamond}_k)\partial_k[\order{\mathbf{g}}{2}^{\bullet}\delta_{ij} + \order{\mathbf{g}}{2}^{\dagger}_{ij}] + \partial_i(\partial_k\order{X}{2}^{\bdiamond} + \order{X}{2}^{\diamond}_k)\partial_j(\partial_k\order{X}{2}^{\bdiamond} + \order{X}{2}^{\diamond}_k)\nonumber\\
&\phantom{=}+ \frac{1}{2}\partial_i[(\partial_k\order{X}{2}^{\bdiamond} + \order{X}{2}^{\diamond}_k)\partial_k(\partial_j\order{X}{2}^{\bdiamond} + \order{X}{2}^{\diamond}_j)] + \frac{1}{2}\partial_j[(\partial_k\order{X}{2}^{\bdiamond} + \order{X}{2}^{\diamond}_k)\partial_k(\partial_i\order{X}{2}^{\bdiamond} + \order{X}{2}^{\diamond}_i)]\,,\label{eq:trametric4ij}
\end{align}
\end{subequations}
where we have written the gauge defining vector fields in the form \(\order{X}{k} = X_{(k)}\epsilon^k/k!\), before using a decomposition
\begin{equation}\label{eq:vecfielddecomp}
X_i = \partial_iX^{\bdiamond} + X^{\diamond}_i\,, \quad
X_0 = X^{\star}\,,
\end{equation}
where \(\partial^iX^{\diamond}_i = 0\). Recalling our definition of the gauge-invariant quantities in section~\ref{ssec:ginvdef}, we remark that the distinguished gauge \(\mathcal{S}\), in which the metric reduces to the gauge-invariant form, is given by the choice \(\order{S}{k} = 0\) for the gauge defining vector fields at all orders.

The gauge transformation~\eqref{eq:trametric} now also clarifies the split of the metric \(\gauge{g}{X}\) into the gauge-invariant part \(\mathbf{g}\) and the choice of the gauge which we mentioned in section~\ref{ssec:ginvdef}:
\begin{enumerate}
\item
At the second velocity order, we see that \(\gaor{g}{X}{2}_{ij}\) contains in addition to the trace part \(\order{\mathbf{g}}{2}^{\bullet}\) and the transverse, trace-free part \(\order{\mathbf{g}}{2}^{\dagger}_{ij}\) also an off-diagonal second derivative and the derivative of a divergence-free vector obtained from the gauge transformation vector field components \(\order{X}{2}^{\bdiamond}\) and \(\order{X}{2}^{\diamond}_i\). The former two quantities have \(1 + 2 = 3\) free components, while the latter have another \(1 + 2 = 3\) free components. Their sum thus equals the six free components of a symmetric tensor of rank two in three dimensions.

\item
Continuing with the third velocity order, we see that the three components of \(\gaor{g}{X}{3}_{0i}\) split into the two free components of the divergence-free vector \(\order{\mathbf{g}}{3}^{\diamond}_i\) and one component for the pure divergence of \(\order{X}{3}^{\star}\). Note that the components of the second-order gauge vector field have already been counted at the previous velocity order and must not be counted again.

\item
The fourth velocity order again yields three free components of the gauge-invariant perturbations and three components of the gauge defining vector fields, as for the second order.
\end{enumerate}
It is the virtue of the gauge-invariant formalism that, once the distinguished gauge \(\mathcal{S}\) is fixed, this splitting of the components of the metric in a general gauge \(\mathcal{X}\) is unique. We will make use of this fact in section~\ref{ssec:standardppn}, when we split the standard PPN metric into its gauge invariant part and the gauge choice.

\subsection{Energy-momentum tensor}\label{ssec:enmomtens}
In order to solve the gravitational field equations, we also need a gauge-invariant description of the energy-momentum tensor, which acts as the source of the gravitational field. In the PPN formalism one assumes that the source matter is given by a perfect fluid, whose energy-momentum tensor takes the form
\begin{subequations}\label{eq:gauenmom}
\begin{align}
\gauge{T}{X}_{00} &= \gauge{\rho}{X}\left(1 - \gaor{g}{X}{2}_{00} + (\gauge{v}{X})^2 + \gauge{\Pi}{X}\right) + \mathcal{O}(6)\,,\\
\gauge{T}{X}_{0i} &= -\gauge{\rho}{X}\gauge{v}{X}_i + \mathcal{O}(5)\,,\\
\gauge{T}{X}_{ij} &= \gauge{\rho}{X}\gauge{v}{X}_i\gauge{v}{X}_j + \gauge{p}{X}\delta_{ij} + \mathcal{O}(6)
\end{align}
\end{subequations}
in an arbitrary gauge \(\mathcal{X}\), up to the relevant perturbation order, where one assigns \(\gauge{\rho}{X} \sim \gauge{\Pi}{X} \sim \mathcal{O}(2)\) and \(\gauge{p}{X} \sim \mathcal{O}(4)\). It follows from this assignment that the lowest terms in the perturbative expansion are already of second velocity order \(\mathcal{O}(2)\) for \(\gauge{T}{X}_{00}\), third order \(\mathcal{O}(3)\) for \(\gauge{T}{X}_{0i}\) and fourth order \(\mathcal{O}(4)\) for \(\gauge{T}{X}_{ij}\). Under a gauge transformation defined by the vector fields \(\order{\xi}{k}\) therefore most of the expansion coefficients \(\gaor{T}{X}{k}\) retain their forms, with the complete set of transformation rules given by
\begin{equation}\label{eq:enmomgatra}
\gaor{T}{Y}{2}_{00} = \gaor{T}{X}{2}_{00}\,, \quad
\gaor{T}{Y}{2}_{ij} = \gaor{T}{X}{2}_{ij} = 0\,, \quad
\gaor{T}{Y}{3}_{0i} = \gaor{T}{X}{3}_{0i}\,, \quad
\gaor{T}{Y}{4}_{00} = \gaor{T}{X}{4}_{00} + \order{\xi}{2}_i\partial_i\gaor{T}{X}{2}_{00}\,, \quad
\gaor{T}{Y}{4}_{ij} = \gaor{T}{X}{4}_{ij}\,.
\end{equation}
To obtain a gauge-invariant expression, we replace all tensors occurring in the formulas above by their gauge invariant counterparts, and introduce a decomposition given by
\begin{subequations}\label{eq:gienmomtens}
\begin{align}
\mathbf{T}^{\star} &= \mathbf{T}_{00} = \boldsymbol{\rho}\left(1 - \order{\mathbf{g}}{2}_{00} + \mathbf{v}^2 + \boldsymbol{\Pi}\right) + \mathcal{O}(6)\,,\\
\mathbf{T}^{\diamond}_i + \partial_i\mathbf{T}^{\bdiamond} &= \mathbf{T}_{0i} = -\boldsymbol{\rho}\mathbf{v}_i + \mathcal{O}(5)\,,\\
\mathbf{T}^{\bullet}\delta_{ij} + \triangle_{ij}\mathbf{T}^{\blacktriangle} + 2\partial_{(i}\mathbf{T}^{\triangle}_{j)} + \mathbf{T}^{\dagger}_{ij} &= \mathbf{T}_{ij} = \boldsymbol{\rho}\mathbf{v}_i\mathbf{v}_j + \mathbf{p}\delta_{ij} + \mathcal{O}(6)\,.\label{eq:gienmomtenst}
\end{align}
\end{subequations}
The terms on the left hand side are the gauge-invariant potentials we are looking for. As it is also the case for the metric, we impose a number of restrictions on these potentials, which are given by
\begin{equation}
\partial^i\mathbf{T}^{\diamond}_i = 0\,, \quad
\partial^i\mathbf{T}^{\triangle}_i = 0\,, \quad
\partial^i\mathbf{T}^{\dagger}_{ij} = 0\,, \quad
\mathbf{T}^{\dagger}_{[ij]} = 0\,, \quad
\mathbf{T}^{\dagger}_{ii} = 0\,.
\end{equation}
Further, we have introduced the notation \(\triangle_{ij} = \partial_i\partial_j - \frac{1}{3}\delta_{ij}\triangle\) for the trace-free second derivative, where \(\triangle = \partial^i\partial_i\) is the spatial Laplace operator. Using the gauge-invariant expressions~\eqref{eq:gienmomtens}, as well as the transformation rules~\eqref{eq:enmomgatra}, we find that in an arbitrary gauge \(\mathcal{X}\) the velocity orders of the energy-momentum tensor are expanded as
\begin{subequations}\label{eq:traenmom}
\begin{align}
\gaor{T}{X}{2}_{00} &= \order{\mathbf{T}}{2}^{\star}\,,\\
\gaor{T}{X}{2}_{ij} &= 0\,,\\
\gaor{T}{X}{3}_{0i} &= \order{\mathbf{T}}{3}^{\diamond}_i + \partial_i\order{\mathbf{T}}{3}^{\bdiamond}\,,\\
\gaor{T}{X}{4}_{00} &= \order{\mathbf{T}}{4}^{\star} + (\partial_i\order{X}{2}^{\bdiamond} + \order{X}{2}^{\diamond}_i)\partial_i\order{\mathbf{T}}{2}^{\star}\,,\\
\gaor{T}{X}{4}_{ij} &= \order{\mathbf{T}}{4}^{\bullet}\delta_{ij} + \triangle_{ij}\order{\mathbf{T}}{4}^{\blacktriangle} + 2\partial_{(i}\order{\mathbf{T}}{4}^{\triangle}_{j)} + \order{\mathbf{T}}{4}^{\dagger}_{ij}\,.
\end{align}
\end{subequations}
These expressions will be useful for solving the field equations of a given gravity theory, as they generically appear on the right hand side of the field equations. We will show this explicitly in section~\ref{sec:example}.

\subsection{Post-Newtonian potentials}\label{ssec:ppnpot}
Another important ingredient of the PPN formalism is the definition of a number of potentials, which are obtained as Poisson-like integrals over the source matter. Note that these integrals, which are defined on the background spacetime \(M_0\) and carried out over a fixed time slice \(t = \text{const}.\), depend on the choice of the coordinates, and hence on the choice of the gauge. In a fixed gauge \(\mathcal{X}\) they are given by the super- and Newtonian potentials at the second velocity order
\begin{equation}
\gauge{\chi}{X}(t,\vec{x}) = -\int d^3x'\gauge{\rho}{X}(t,\vec{x}')|\vec{x} - \vec{x}'|\,, \quad
\gauge{U}{X}(t,\vec{x}) = \int d^3x'\frac{\gauge{\rho}{X}(t,\vec{x}')}{|\vec{x} - \vec{x}'|}\,,
\end{equation}
the third-order vector potentials
\begin{equation}
\gauge{V}{X}_i(t,\vec{x}) = \int d^3x'\frac{\gauge{\rho}{X}(t,\vec{x}')\gauge{v}{X}_i(t,\vec{x}')}{|\vec{x} - \vec{x}'|}\,,\quad
\gauge{W}{X}_i(t,\vec{x}) = \int d^3x'\frac{\gauge{\rho}{X}(t,\vec{x}')\gauge{v}{X}_j(t,\vec{x}')(x_i - x_i')(x_j - x_j')}{|\vec{x} - \vec{x}'|^3}\,,
\end{equation}
as well as the fourth-order scalar potentials
\begin{gather}
\gauge{\Phi}{X}_1(t,\vec{x}) = \int d^3x'\frac{\gauge{\rho}{X}(t,\vec{x}')\gauge{v}{X}^2(t,\vec{x}')}{|\vec{x} - \vec{x}'|}\,,\quad
\gauge{\Phi}{X}_2(t,\vec{x}) = \int d^3x'\frac{\gauge{\rho}{X}(t,\vec{x}')\gauge{U}{X}(t,\vec{x}')}{|\vec{x} - \vec{x}'|}\,,\nonumber\\
\gauge{\Phi}{X}_3(t,\vec{x}) = \int d^3x'\frac{\gauge{\rho}{X}(t,\vec{x}')\gauge{\Pi}{X}(t,\vec{x}')}{|\vec{x} - \vec{x}'|}\,,\quad
\gauge{\Phi}{X}_4(t,\vec{x}) = \int d^3x'\frac{\gauge{p}{X}(t,\vec{x}')}{|\vec{x} - \vec{x}'|}\,,\nonumber\\
\gauge{\mathfrak{A}}{X}(t,\vec{x}) = \int d^3x'\frac{\gauge{\rho}{X}(t,\vec{x}')\left[\gauge{v}{X}_i(t,\vec{x}')(x_i - x_i')\right]^2}{|\vec{x} - \vec{x}'|^3}\,,\quad
\gauge{\mathfrak{B}}{X}(t,\vec{x}) = \int d^3x'\frac{\gauge{\rho}{X}(t,\vec{x}')}{|\vec{x} - \vec{x}'|}(x_i - x_i')\frac{d\gauge{v}{X}_i(t,\vec{x}')}{dt}\,.\nonumber\\
\gauge{\Phi}{X}_W(t,\vec{x}) = \int d^3x'd^3x''\gauge{\rho}{X}(t,\vec{x}')\gauge{\rho}{X}(t,\vec{x}'')\frac{x_i - x_i'}{|\vec{x} - \vec{x}'|^3}\left(\frac{x_i' - x_i''}{|\vec{x} - \vec{x}''|} - \frac{x_i - x_i''}{|\vec{x}' - \vec{x}''|}\right)\,.
\end{gather}
In a different gauge \(\mathcal{Y}\) the PPN potentials are defined analogously with \(\mathcal{Y}\) in place of \(\mathcal{X}\). To relate the PPN potentials in the different gauges, one may simply transform the corresponding quantities which appear inside the integrals and which define the fluid source matter. In particular, using the transformation
\begin{equation}\label{eq:rhotrans}
\gauge{\rho}{Y} = \gauge{\rho}{X} + \order{\xi}{2}_i\gauge{\rho}{X}_{,i} + \mathcal{O}(6)\,,
\end{equation}
obtained from the corresponding Lie derivative, we find that the second-order potentials transform as
\begin{subequations}
\begin{align}
\gauge{\chi}{Y}(t,\vec{x}) &= \gauge{\chi}{X}(t,\vec{x}) + \int d^3x'\gauge{\rho}{X}(t,\vec{x}')\left(\partial'_i\order{\xi}{2}_i(t,\vec{x}')|\vec{x} - \vec{x}'| - \frac{\order{\xi}{2}_i(t,\vec{x}')(x_i - x'_i)}{|\vec{x} - \vec{x}'|}\right) + \mathcal{O}(6)\,,\label{eq:trapotchi}\\
\gauge{U}{Y}(t,\vec{x}) &= \gauge{U}{X}(t,\vec{x}) - \int d^3x'\gauge{\rho}{X}(t,\vec{x}')\left(\frac{\partial'_i\order{\xi}{2}_i(t,\vec{x}')}{|\vec{x} - \vec{x}'|} + \frac{\order{\xi}{2}_i(t,\vec{x}')(x_i - x'_i)}{|\vec{x} - \vec{x}'|^3}\right) + \mathcal{O}(6)\,.\label{eq:trapotu}
\end{align}
\end{subequations}
The simplicity of deriving these relations, using only the formula~\eqref{eq:rhotrans}, shows one of the advantages of our geometric interpretation of gauge compared to the conventional interpretation in terms of coordinate choices. In our formulation, the coordinates stay fixed, and only the source field must be transformed, i.e., pulled back using a different diffeomorphism. The conventional formulation involves a coordinate transformation instead, which enters the integrals in multiple places due to their explicit coordinate dependence, so that the derivation becomes more lengthy and requires particular care given to all appearing terms. The result is, of course, the same~\cite{Will:1993ns}. Note that similar calculations show the transformation behavior of the remaining potentials. However, we will not discuss their transformations explicitly, since any arising differences are of negligible perturbation order, so that they will not be necessary for our following treatment.

Finally, we may now use the PPN potentials in the distinguished gauge \(\mathcal{S}\), which we write in boldface in analogy to the other quantities in this gauge, to express the gauge-invariant parts of the energy-momentum tensor~\eqref{eq:gienmomtens}. By taking appropriate traces and divergences, and solving the resulting Poisson-like equations, we obtain the relations
\begin{subequations}\label{eq:enmomppnpot}
\begin{align}
\order{\mathbf{T}}{2}^{\star} &= \boldsymbol{\rho} = -\frac{1}{4\pi}\triangle\mathbf{U}\,,\\
\order{\mathbf{T}}{3}^{\bdiamond} &= -\frac{1}{4\pi}\partial_0\mathbf{U}\,,\\
\order{\mathbf{T}}{3}^{\diamond}_i &= \frac{1}{8\pi}\triangle(\mathbf{V}_i + \mathbf{W}_i)\,,\\
\order{\mathbf{T}}{4}^{\star} &= \boldsymbol{\rho}\left(\boldsymbol{\Pi} + \mathbf{v}^2 - \order{\mathbf{g}}{2}^{\star}\right) = -\frac{1}{4\pi}\triangle\left(\boldsymbol{\Phi}_3 + \boldsymbol{\Phi}_1 - 2\boldsymbol{\Phi}_2\right)\,,\label{eq:enmomnewtlim}\\
\order{\mathbf{T}}{4}^{\bullet} &= \frac{1}{3}\boldsymbol{\rho}\mathbf{v}^2 + \mathbf{p} = -\frac{1}{12\pi}\triangle(\boldsymbol{\Phi}_1 + 3\boldsymbol{\Phi}_4)\,,\\
\order{\mathbf{T}}{4}^{\blacktriangle} &= \frac{1}{16\pi}(3\boldsymbol{\mathfrak{A}} - \boldsymbol{\Phi}_1)\,.
\end{align}
\end{subequations}
Note that for the component~\eqref{eq:enmomnewtlim} we have already made use of the Newtonian limit, which mandates that the component \(\order{\mathbf{g}}{2}^{\star}\) must be given by \(\order{\mathbf{g}}{2}^{\star} = 2\mathbf{U}\); this will become more clear in the next section. Further, we have omitted the fourth order vector and tensor components \(\order{\mathbf{T}}{4}^{\triangle}_i\) and \(\order{\mathbf{T}}{4}^{\dagger}_{ij}\), since at the fourth velocity order generally only the scalar components appear in the field equations for the scalar components of the fourth order metric we need to solve for. However, if needed one may derive these quantities by taking the divergence of their defining relation~\eqref{eq:gienmomtenst} and integrating the resulting Poisson-like equation. The result can be brought into different, equivalent forms by using the Euler equations for the perfect fluid; we do not display these here for brevity.

\subsection{Standard post-Newtonian gauge and PPN parameters}\label{ssec:standardppn}
The final aim of applying the PPN formalism to a theory of gravity is to calculate the so-called PPN parameters. These are defined as the (constant) coefficients which appear if one expresses the post-Newtonian metric, in a specific gauge, through a linear combination of post-Newtonian potentials. This specific gauge, which we denote \(\mathcal{P}\), is called the standard post-Newtonian gauge. In this gauge, the metric is assumed to be of the form~\cite{Will:1993ns}
\begin{subequations}\label{eq:standardppn}
\begin{align}
\gaor{g}{P}{2}_{00} &= 2\gauge{U}{P}\,,\label{eq:standardppn200}\\
\gaor{g}{P}{2}_{ij} &= 2\gamma\gauge{U}{P}\delta_{ij}\,,\label{eq:standardppn2ij}\\
\gaor{g}{P}{3}_{0i} &= -\frac{1}{2}(3 + 4\gamma + \alpha_1 - \alpha_2 + \zeta_1 - 2\xi)\gauge{V}{P}_i - \frac{1}{2}(1 + \alpha_2 - \zeta_1 + 2\xi)\gauge{W}{P}_i\,,\label{eq:standardppn30i}\\
\gaor{g}{P}{4}_{00} &= -2\beta\gauge{U}{P}^2 + (2 + 2\gamma + \alpha_3 + \zeta_1 - 2\xi)\gauge{\Phi}{P}_1 + 2(1 + 3\gamma - 2\beta + \zeta_2 + \xi)\gauge{\Phi}{P}_2\nonumber\\
&\phantom{=}+ 2(1 + \zeta_3)\gauge{\Phi}{P}_3 + 2(3\gamma + 3\zeta_4 - 2\xi)\gauge{\Phi}{P}_4 - 2\xi\gauge{\Phi}{P}_W - (\zeta_1 - 2\xi)\gauge{\mathfrak{A}}{P}\,.\label{eq:standardppn400}
\end{align}
\end{subequations}
The first equation~\eqref{eq:standardppn200} implements the Newtonian limit, where the Newtonian constant is tautologically assumed to be constant and normalized to \(1\). Observe that \(\gaor{g}{P}{2}_{ij}\) is diagonal and that \(\gaor{g}{P}{4}_{00}\) does not contain the potential \(\gauge{\mathfrak{B}}{P}\); this is the defining property of the standard PPN gauge. While the former condition is easily implemented in the gauge-invariant formalism we developed here, the latter is rather cumbersome. To see this, note that \(\gaor{g}{P}{3}_{0i}\) contains a non-vanishing divergence part proportional to \(\gauge{V}{P}_i - \gauge{W}{P}_i\), which is determined only after fixing the component \(\order{P}{3}^{\star}\) of the third order gauge vector field. However, this component is fixed only by the absence of \(\gauge{\mathfrak{B}}{P}\) from \(\gaor{g}{P}{4}_{00}\), and so the third order metric is fully determined only after solving also the fourth order. The latter is significantly more involved than solving the third order equations only, since it requires an expansion of the field equations quadratic in the perturbations, which does not occur in the third order. In the proposed gauge invariant formalism we have avoided this issue by choosing a different gauge condition.

We will now decompose the standard PPN metric~\eqref{eq:standardppn} into its gauge-invariant and gauge dependent parts, i.e., the vector fields \(P\) defining the gauge \(\mathcal{P}\), using the decomposition~\eqref{eq:trametric}. For this purpose, we compare the metrics~\eqref{eq:standardppn} and~\eqref{eq:trametric} at each perturbation order, starting from the lowest. By comparing the second-order temporal components~\eqref{eq:trametric200} and~\eqref{eq:standardppn200} corresponding to the Newtonian limit, and further using the fact that under the gauge transformation~\eqref{eq:trapotu} the Newtonian potential changes only by a term of fourth velocity order \(\mathcal{O}(4)\), so that \(\gauge{U}{P} = \mathbf{U} + \mathcal{O}(4)\), one immediately reads off the relation \(\order{\mathbf{g}}{2}^{\star} = 2\mathbf{U}\) for the second velocity order; the fourth order correction term which we obtain here by changing the post-Newtonian potential becomes part of \(\order{\mathbf{g}}{4}^{\star}\) later. For the next step, note that the second-order spatial components of the standard PPN metric~\eqref{eq:standardppn2ij} are assumed to be diagonal. Hence, by comparison with their expansion~\eqref{eq:trametric2ij}, one finds that they uniquely decompose into the components
\begin{equation}\label{eq:ppntra2}
\order{\mathbf{g}}{2}^{\bullet} = 2\gamma\mathbf{U}\,, \quad
\order{\mathbf{g}}{2}^{\dagger}_{ij} = 0\,, \quad
\order{P}{2}^{\bdiamond} = 0\,, \quad
\order{P}{2}^{\diamond}_i = 0\,,
\end{equation}
since only the first of these contains the diagonal (trace) part, while the remaining three components would introduce off-diagonal terms. One then continues with the third velocity order. Here the decomposition~\eqref{eq:trametric30i} mandates to separate the pure divergence \(\partial_i\order{P}{3}^{\star} + \partial_0\partial_i\order{P}{2}^{\bdiamond}\) from the divergence-free part \(\order{\mathbf{g}}{3}^{\diamond}_i + \partial_0\order{P}{2}^{\diamond}_i\). To apply this decomposition to the standard PPN metric~\eqref{eq:standardppn30i}, first note that the third-order PPN potentials transform under a gauge transformation as \(\gauge{V}{P}_i = \mathbf{V}_i + \mathcal{O}(5)\) and \(\gauge{W}{P}_i = \mathbf{W}_i + \mathcal{O}(5)\), so that we can immediately work with the gauge-invariant potentials and neglect the higher-order contribution. From their definition one can derive that they satisfy the relations~\cite{Will:1993ns}
\begin{equation}
\partial_i\mathbf{V}_i = -\partial_i\mathbf{W}_i = -\partial_0\mathbf{U}\,, \quad
\mathbf{V}_i - \mathbf{W}_i = \partial_0\partial_i\boldsymbol{\chi}\,.
\end{equation}
The former shows that their sum \(\mathbf{V}_i + \mathbf{W}_i\) is divergence-free, while their difference \(\mathbf{V}_i - \mathbf{W}_i\) is a pure divergence. Hence, the unique decomposition~\eqref{eq:trametric30i} of the component~\eqref{eq:standardppn30i} yields the gauge dependent part
\begin{equation}\label{eq:ppngvecf}
\order{P}{3}^{\star} = -\frac{1}{4}(2 + 4\gamma + \alpha_1 - 2\alpha_2 + 2\zeta_1 - 4\xi)\boldsymbol{\chi}_{,0}\,,
\end{equation}
as well as the gauge-invariant part~\eqref{eq:gimet3} which we display below. Finally, inserting the component~\eqref{eq:standardppn400} and the gauge defining vector field~\eqref{eq:ppngvecf} into the decomposition~\eqref{eq:trametric400}, we can solve for the gauge-invariant component \(\order{\mathbf{g}}{4}^{\star}\). Here we must take into account that at the second velocity order component \(\order{\mathbf{g}}{2}^{\star}\) we replaced the Newtonian potential \(\gauge{U}{P}\) by its gauge-invariant counterpart \(\mathbf{U}\). Their difference, which is of fourth velocity order \(\mathcal{O}(4)\), must therefore be taken into account as contribution to \(\order{\mathbf{g}}{4}^{\star}\). However, note that from the second order equations~\eqref{eq:ppntra2} follows that the second-order gauge defining vector fields vanish, \(\order{P}{2}^{\bdiamond} = 0\) and \(\order{P}{2}^{\diamond}_i = 0\). Together with the transformation rule~\eqref{eq:trapotu} this implies that also the aforementioned fourth-order correction vanishes. This yields the fourth-order gauge-invariant component~\eqref{eq:gimet4}, so that we may summarize the full list of gauge-invariant components determined by the PPN metric as
\begin{subequations}\label{eq:ginvppnmet}
\begin{align}
\order{\mathbf{g}}{2}^{\star} &= 2\mathbf{U}\label{eq:ginewtlim}\,,\\
\order{\mathbf{g}}{2}^{\bullet} &= 2\gamma\mathbf{U}\,,\\
\order{\mathbf{g}}{2}^{\dagger}_{ij} &= 0\,,\\
\order{\mathbf{g}}{3}^{\diamond}_i &= -\left(1 + \gamma + \frac{\alpha_1}{4}\right)(\mathbf{V}_i + \mathbf{W}_i)\,,\label{eq:gimet3}\\
\order{\mathbf{g}}{4}^{\star} &= \frac{1}{2}(2 - \alpha_1 + 2\alpha_2 + 2\alpha_3)\boldsymbol{\Phi}_1 + 2(1 + 3\gamma - 2\beta + \zeta_2 + \xi)\boldsymbol{\Phi}_2 + 2(1 + \zeta_3)\boldsymbol{\Phi}_3 + 2(3\gamma + 3\zeta_4 - 2\xi)\boldsymbol{\Phi}_4\nonumber\\
&\phantom{=}- 2\xi\boldsymbol{\Phi}_W - 2\beta\mathbf{U}^2 + \frac{1}{2}(2 + 4\gamma + \alpha_1 - 2\alpha_2)\boldsymbol{\mathfrak{A}} + \frac{1}{2}(2 + 4\gamma + \alpha_1 - 2\alpha_2 + 2\zeta_1 - 4\xi)\boldsymbol{\mathfrak{B}}\,,\label{eq:gimet4}
\end{align}
\end{subequations}
where in the last component we used another relation following from the definition of the PPN potentials given by~\cite{Will:1993ns}
\begin{equation}
\boldsymbol{\chi}_{,00} = \boldsymbol{\mathfrak{A}} + \boldsymbol{\mathfrak{B}} - \boldsymbol{\Phi}_1\,.
\end{equation}
A few remarks are in order. First note that the relation~\eqref{eq:ginewtlim} is the gauge-invariant formulation of the Newtonian limit~\eqref{eq:standardppn200}. Second, in the standard PPN metric the tensor part \(\order{\mathbf{g}}{2}^{\dagger}_{ij}\) is assumed to vanish identically. This assumption is related to the fact that at the second velocity order the only source term arising from the energy-momentum tensor of a perfect fluid is a scalar, and so there is neither a source term for \(\order{\mathbf{g}}{2}^{\dagger}_{ij}\), nor a possibility to construct a corresponding PPN potential. However, if one considers more general source matter models exhibiting also anisotropic stress, one may suitably extend the PPN formalism to include also such terms in the PPN metric.

Further, we see that already by solving the field equations up to the second velocity order we obtain the PPN parameter \(\gamma\), while solving for the third velocity order also yields the PPN parameter \(\alpha_1\). This is less obvious in the standard PPN metric~\eqref{eq:standardppn}, where the final form of \(\gaor{g}{P}{3}_{0i}\) is determined only after gauge fixing by eliminating \(\gauge{\mathfrak{B}}{P}\) from (and hence solving for) \(\gaor{g}{P}{4}_{00}\). However, note that of course also in the standard formalism one can obtain \(\alpha_1\) already after solving the field equations at the third velocity order, by decomposing \(\gaor{g}{P}{3}_{0i}\) into its pure divergence and divergence-free parts, as explained in section~\ref{ssec:ginvmetric}, hence effectively calculating \(\order{\mathbf{g}}{3}^{\diamond}_i\).

We also remark that the standard PPN gauge \(\mathcal{P}\) is not related to the field equations of any particular gravity theory. In the context of gauge-invariant perturbations we discuss here, it is obtained from the generic linear combination~\eqref{eq:ginvppnmet} of the PPN potentials in the gauge-invariant metric components by applying the unique gauge transformation defined by the second-order vector fields~\eqref{eq:ppntra2} which retains the diagonal form of \(\gaor{g}{P}{2}_{ij}\) and the third-order vector field~\eqref{eq:ppngvecf} which cancels the potential \(\boldsymbol{\mathfrak{B}}\) from the fourth-order component \(\gaor{g}{P}{4}_{00}\), as shown explicitly in~\cite[Sec. 4.2]{Will:1993ns}.

Note that one could easily extend the post-Newtonian metric~\eqref{eq:ginvppnmet} beyond the standard PPN formalism by including, e.g., an expansion of the component \(\order{\mathbf{g}}{4}^{\bullet}\) in terms of PPN potentials, thereby introducing new PPN parameters as their coefficients. While this is not done in the standard PPN formalism, since it yields a subleading contribution to the motion of slow-moving test matter, it may be used, e.g., to calculate the deflection of light at higher post-Newtonian orders~\cite{Richter:1982zz,Richter:1982zza}, as argued at the end of section~\ref{ssec:ppnpert}. We will not pursue this direction here, as it would go beyond the aim of this article.

This completes our construction of a gauge-invariant PPN formalism. We will demonstrate its use by applying it to an example theory in section~\ref{sec:example}. However, before doing so, we also present a tetrad formulation in the following section.

\section{Tetrad formulation}\label{sec:tetrad}
In the previous section we have derived a gauge-invariant approach to the PPN formalism in its standard, metric formulation. While this is suitable for most theories of gravity, there are also theories which employ a tetrad instead of the metric as their fundamental field variable. We therefore also present a tetrad formulation of our formalism in this section. We proceed in analogy to the metric case. In section~\ref{ssec:tetpert} we discuss the perturbative expansion of the tetrad in velocity orders and its relation to the corresponding expansion of the metric. Gauge transformations of the tetrad perturbations are discussed in section~\ref{ssec:tetgtrans}. These are then used in section~\ref{ssec:ginvtet} to derive a gauge-invariant set of tetrad perturbations. Finally, in section~\ref{ssec:tetppn} we relate the gauge invariant tetrad perturbations to the standard PPN gauge and thus the PPN parameters.

\subsection{Perturbative expansion}\label{ssec:tetpert}
The starting point for our tetrad formulation of the gauge-invariant PPN formalism is a perturbative expansion of the tetrad around a fixed background, in analogy to the expansion of the metric shown in section~\ref{ssec:ppnpert}. For this purpose, note first that the metric \(g_{\mu\nu}\) and the tetrad \(\theta^A{}_{\mu}\) are related by
\begin{equation}\label{eq:tettomet}
g_{\mu\nu} = \eta_{AB}\theta^A{}_{\mu}\theta^B{}_{\nu}\,,
\end{equation}
where \(\eta_{AB} = \mathrm{diag}(-1, 1, 1, 1)\) is the Minkowski metric and we introduced Lorentz indices \(A, B = 0, \ldots, 3\). In a gauge \(\mathcal{X}\), we then write the tetrad in a perturbative expansion of the form
\begin{equation}\label{eq:tetppnexp}
\gauge{\theta}{X} = \sum_{k = 0}^{\infty}\gaor{\theta}{X}{k}\,.
\end{equation}
To be consistent with the corresponding expansion~\eqref{eq:metppnexp} we set the zeroth order term to the diagonal tetrad, \(\gaor{\theta}{X}{0}^A{}_{\mu} = \Delta^A{}_{\mu} = \mathrm{diag}(1, 1, 1, 1)\), which defines the background geometry. For the remaining terms \(k > 0\) it then turns out to be more convenient to lower the Lorentz index with the Minkowski metric and to turn it into a spacetime index with the diagonal tetrad, so that we define
\begin{equation}\label{eq:tetlowdef}
\gaor{\theta}{X}{k}_{\mu\nu} = \Delta^A{}_{\mu}\eta_{AB}\gaor{\theta}{X}{k}^B{}_{\nu}\,.
\end{equation}
In order to restrict the tetrad perturbation components we consider, we follow a similar line of arguments as for the metric treatment at the end of section~\ref{ssec:ppnpert}., one can show that only certain components of the tetrad perturbations are relevant in the PPN formalism. In analogy to the metric perturbation components~\eqref{eq:metcompgauge}, these are given by
\begin{equation}\label{eq:tetcompgauge}
\gaor{\theta}{X}{2}_{00}\,, \quad
\gaor{\theta}{X}{2}_{ij}\,, \quad
\gaor{\theta}{X}{3}_{0i}\,, \quad
\gaor{\theta}{X}{3}_{i0}\,, \quad
\gaor{\theta}{X}{4}_{00}\,, \quad
\gaor{\theta}{X}{4}_{ij}\,.
\end{equation}
Note that also here the last component \(\gaor{\theta}{X}{4}_{ij}\) would not appear in a naive treatment of the standard PPN formalism, but we keep it here, since it will in general be coupled to the component \(\gaor{\theta}{X}{4}_{00}\) we must solve for to determine the PPN parameters. This can also be seen by deriving the relation between the metric perturbations~\eqref{eq:metppnexp} and the tetrad perturbations~\eqref{eq:tetppnexp}. Expanding the relation~\eqref{eq:tettomet} in velocity orders in a given gauge yields
\begin{equation}\label{eq:metrictetrad}
\gaor{g}{X}{2}_{00} = 2\gaor{\theta}{X}{2}_{00}\,, \quad
\gaor{g}{X}{2}_{ij} = 2\gaor{\theta}{X}{2}_{(ij)}\,, \quad
\gaor{g}{X}{3}_{0i} = 2\gaor{\theta}{X}{3}_{(0i)}\,, \quad
\gaor{g}{X}{4}_{00} = -(\gaor{\theta}{X}{2}_{00})^2 + 2\gaor{\theta}{X}{4}_{00}\,, \quad
\gaor{g}{X}{4}_{ij} = 2\gaor{\theta}{X}{4}_{(ij)} + \gaor{\theta}{X}{2}_{ki}\gaor{\theta}{X}{2}_{kj}\,.
\end{equation}
Here we also see that only the symmetric parts of the tetrads are relevant for the metric perturbations, with the exception of \(\gaor{\theta}{X}{2}_{ij}\), where also the antisymmetric part enters in the last term. It will thus be useful to treat the symmetric and antisymmetric parts separately, and to define
\begin{equation}
\gaor{\theta}{X}{k}_{\mu\nu} = \gaor{s}{X}{k}_{\mu\nu} + \gaor{a}{X}{k}_{\mu\nu}\,, \quad
\gaor{s}{X}{k}_{\mu\nu} = \gaor{\theta}{X}{k}_{(\mu\nu)}\,, \quad
\gaor{a}{X}{k}_{\mu\nu} = \gaor{\theta}{X}{k}_{[\mu\nu]}\,.
\end{equation}
In the following we will make use of this decomposition.

\subsection{Gauge transformations}\label{ssec:tetgtrans}
Using the perturbative expansions given above, one can now proceed similarly to the derivation in section~\ref{ssec:gtrans} and calculate the gauge transformation of the tetrad perturbations. These follow from the same formula~\eqref{eq:pertgtrans} which holds also for the metric perturbations. However, when applying this formula, one must pay attention that the tetrads are one-forms with an additional Lorentz index, which comes from the fact that they take values in a vector bundle. Note that there is in general no a priori relation between these Lorentz vector bundles over the physical spacetime \(M\) and the background spacetime \(M_0\). Hence, the full set of gauge transformations for a tetrad is given not only by diffeomorphisms relating \(M\) and \(M_0\), but by vector bundle isomorphisms relating these two Lorentz vector bundles, in order to take into account the additional gauge freedom.

In the following we will resort to a simplified treatment, and assume that both on the physical spacetime \(M\) and the background spacetime \(M_0\) a fixed Lorentz gauge for the tetrad is chosen\footnote{An example for such a fixed Lorentz gauge choice used in a tetrad extension of the PPN formalism is the Weitzenböck gauge in teleparallel gravity~\cite{Ualikhanova:2019ygl} and its scalar extension~\cite{Emtsova:2019qsl,Flathmann:2019khc}, which may be imposed independently at each order of the perturbation theory.}, and that the Lorentz index of the tetrad reflects its component with respect to this fixed gauge. As a consequence, the Lorentz index is inert under gauge transformations, which are again given by diffeomorphisms as in the metric case, so that the Lie derivative acts only on the spacetime index. This means that one must use the index form \(\theta^A{}_{\mu}\) for the tetrads, and not the transformed index expression~\eqref{eq:tetlowdef}, which can only be used by properly taking into account also the Lie derivative of the background tetrad \(\Delta^A{}_{\mu}\). First applying the gauge transformation and then transforming the indices using the relation~\eqref{eq:tetlowdef} then yields the transformation behavior
\begin{subequations}\label{eq:ppntetgatra}
\begin{align}
\gaor{\theta}{Y}{2}_{00} &= \gaor{\theta}{X}{2}_{00}\,,\\
\gaor{\theta}{Y}{2}_{ij} &= \gaor{\theta}{X}{2}_{ij} + \partial_j\order{\xi}{2}_i\,,\\
\gaor{\theta}{Y}{3}_{0i} &= \gaor{\theta}{X}{3}_{0i} + \partial_i\order{\xi}{3}_0\,,\\
\gaor{\theta}{Y}{3}_{i0} &= \gaor{\theta}{X}{3}_{i0} + \partial_0\order{\xi}{2}_i\,,\\
\gaor{\theta}{Y}{4}_{00} &= \gaor{\theta}{X}{4}_{00} + \partial_0\order{\xi}{3}_0 + \order{\xi}{2}_i\partial_i\gaor{\theta}{X}{2}_{00}\,,\\
\gaor{\theta}{Y}{4}_{ij} &= \gaor{\theta}{X}{4}_{ij} + \partial_j\order{\xi}{4}_i + \partial_j\order{\xi}{2}_k\gaor{\theta}{X}{2}_{ik} + \order{\xi}{2}_k\partial_k\gaor{\theta}{X}{2}_{ij} + \frac{1}{2}\partial_j(\order{\xi}{2}_k\partial_k\order{\xi}{2}_i)\,.
\end{align}
\end{subequations}
As it is also the case for the metric, one can now eliminate certain components of the tetrad perturbations by a suitable choice of the gauge transformation. Choosing the components which are to be eliminated will then allow us to fix the gauge.

\subsection{Gauge-invariant tetrad}\label{ssec:ginvtet}
We can now make use of the gauge transformation~\eqref{eq:ppntetgatra}, in order to find gauge-invariant components for the tetrad perturbations, as we have done for the metric in section~\ref{ssec:ginvmetric}. First note that by a suitable choice of \(\xi_i\) it is possible to eliminate certain components of \(\theta_{ij}\). One possible choice is to retain only a diagonal (pure trace) and a symmetric, trace-free, divergence-free part, as well as an antisymmetric part. Similarly to the metric case, we could then further choose \(\xi_0\) such that any divergence is eliminated from \(\theta_{0i}\), and retain only a divergence-free part. However, note that one cannot perform such a simplification for the component \(\theta_{i0}\), since \(\xi_i\) is already fixed by the previous condition, and so it contains both a pure divergence and a divergence-free part. This means that the symmetric part \(s_{0i}\), which is relevant for determining the metric, and which we thus aim to solve for and simplify, would also retain a divergence part. Thus, choosing the divergence part of \(\theta_{0i}\) such that it cancels the contribution from \(\theta_{i0}\) appears more useful. Using this gauge fixing, we can parametrize the resulting tetrad in the form
\begin{equation}\label{eq:gitetrad}
\boldsymbol{\theta}_{00} = \boldsymbol{\theta}^{\star}\,, \quad
\boldsymbol{\theta}_{0i} = \partial_i\boldsymbol{\theta}^{\bdiamond} + \boldsymbol{\theta}^{\diamond}_i + \boldsymbol{\theta}^{\circ}_i\,, \quad
\boldsymbol{\theta}_{i0} = -\partial_i\boldsymbol{\theta}^{\bdiamond} + \boldsymbol{\theta}^{\diamond}_i - \boldsymbol{\theta}^{\circ}_i\,, \quad
\boldsymbol{\theta}_{ij} = \boldsymbol{\theta}^{\bullet}\delta_{ij} + \boldsymbol{\theta}^{\dagger}_{ij} + \epsilon_{ijk}(\partial_k\boldsymbol{\theta}^{\blacksquare} + \boldsymbol{\theta}^{\square}_k)\,.
\end{equation}
Note that we have chosen a parametrization which simplifies the split of the tetrad perturbations into symmetric and antisymmetric parts. These are given by
\begin{equation}\label{eq:giastetrad}
\mathbf{s}_{00} = \boldsymbol{\theta}^{\star}\,, \quad
\mathbf{s}_{0i} = \boldsymbol{\theta}^{\diamond}_i\,, \quad
\mathbf{s}_{ij} = \boldsymbol{\theta}^{\bullet}\delta_{ij} + \boldsymbol{\theta}^{\dagger}_{ij}\,, \quad
\mathbf{a}_{0i} = \partial_i\boldsymbol{\theta}^{\bdiamond} + \boldsymbol{\theta}^{\circ}_i\,, \quad
\mathbf{a}_{ij} = \epsilon_{ijk}(\partial_k\boldsymbol{\theta}^{\blacksquare} + \boldsymbol{\theta}^{\square}_k)\,.
\end{equation}
We can then transform the tetrad perturbations to an arbitrary gauge \(\mathcal{X}\). Decomposing the gauge defining vector fields in the form~\eqref{eq:vecfielddecomp} as in the metric case, we find the transformation behavior
\begin{subequations}\label{eq:tratetrad}
\begin{align}
\gaor{\theta}{X}{2}_{00} &= \order{\boldsymbol{\theta}}{2}^{\star}\,,\\
\gaor{\theta}{X}{2}_{ij} &= \order{\boldsymbol{\theta}}{2}^{\bullet}\delta_{ij} + \order{\boldsymbol{\theta}}{2}^{\dagger}_{ij} + \epsilon_{ijk}(\partial_k\order{\boldsymbol{\theta}}{2}^{\blacksquare} + \order{\boldsymbol{\theta}}{2}^{\square}_k) + \partial_i\partial_j\order{X}{2}^{\bdiamond} + \partial_j\order{X}{2}^{\diamond}_i\,,\\
\gaor{\theta}{X}{3}_{0i} &= \partial_i\order{\boldsymbol{\theta}}{3}^{\bdiamond} + \order{\boldsymbol{\theta}}{3}^{\diamond}_i + \order{\boldsymbol{\theta}}{3}^{\circ}_i + \partial_i\order{X}{3}^{\star}\,,\\
\gaor{\theta}{X}{3}_{i0} &= -\partial_i\order{\boldsymbol{\theta}}{3}^{\bdiamond} + \order{\boldsymbol{\theta}}{3}^{\diamond}_i - \order{\boldsymbol{\theta}}{3}^{\circ}_i + \partial_0\partial_i\order{X}{2}^{\bdiamond} + \partial_0\order{X}{2}^{\diamond}_i\,,\\
\gaor{\theta}{X}{4}_{00} &= \order{\boldsymbol{\theta}}{4}^{\star} + \partial_0\order{X}{3}^{\star} + \partial_i\order{\boldsymbol{\theta}}{2}^{\star}(\partial_i\order{X}{2}^{\bdiamond} + \order{X}{2}^{\diamond}_i)\,,\\
\gaor{\theta}{X}{4}_{ij} &= \order{\boldsymbol{\theta}}{4}^{\bullet}\delta_{ij} + \order{\boldsymbol{\theta}}{4}^{\dagger}_{ij} + \epsilon_{ijk}(\partial_k\order{\boldsymbol{\theta}}{4}^{\blacksquare} + \order{\boldsymbol{\theta}}{4}^{\square}_k) + \partial_i\partial_j\order{X}{4}^{\bdiamond} + \partial_j\order{X}{4}^{\diamond}_i + \frac{1}{2}\partial_j[(\partial_k\order{X}{2}^{\bdiamond} + \order{X}{2}^{\diamond}_k)\partial_k(\partial_i\order{X}{2}^{\bdiamond} + \order{X}{2}^{\diamond}_i)] \nonumber\\
&\phantom{=}+ (\partial_k\order{X}{2}^{\bdiamond} + \order{X}{2}^{\diamond}_k)\partial_k[\order{\boldsymbol{\theta}}{2}^{\bullet}\delta_{ij} + \order{\boldsymbol{\theta}}{2}^{\dagger}_{ij} + \epsilon_{ijl}(\partial_l\order{\boldsymbol{\theta}}{2}^{\blacksquare} + \order{\boldsymbol{\theta}}{2}^{\square}_l)] + \partial_j(\partial_k\order{X}{2}^{\bdiamond} + \order{X}{2}^{\diamond}_k)[\order{\boldsymbol{\theta}}{2}^{\bullet}\delta_{ik} + \order{\boldsymbol{\theta}}{2}^{\dagger}_{ik} + \epsilon_{ikl}(\partial_l\order{\boldsymbol{\theta}}{2}^{\blacksquare} + \order{\boldsymbol{\theta}}{2}^{\square}_l)]\,.
\end{align}
\end{subequations}
It must be noted that the distinguished gauge in which the tetrad takes the form~\eqref{eq:gitetrad} is not the same as the one we used before to bring the metric to the form~\eqref{eq:gimetric}. This can be seen, for example, by noticing that \(\order{\mathbf{g}}{4}_{ij}\) contains only a pure trace and a symmetric, trace-free, divergence-free part, while the corresponding metric component \(2\order{\boldsymbol{\theta}}{4}_{(ij)} + \order{\boldsymbol{\theta}}{2}_{ki}\order{\boldsymbol{\theta}}{2}_{kj}\) obtained from the gauge-invariant tetrad receives additional contributions from the second, non-linear term. However, since this is the only component in which the different gauge choice appears, and its form is not relevant for determining the PPN parameters, it will not make any difference for our calculations.

\subsection{Standard post-Newtonian gauge and PPN parameters}\label{ssec:tetppn}
We can now finally establish the relation between the gauge-invariant tetrad perturbations and the PPN parameters, in analogy to the relation~\eqref{eq:ginvppnmet} we derived in section~\ref{ssec:standardppn} for the metric perturbations. Expressing the metric~\eqref{eq:standardppn} in the standard PPN gauge through the tetrad by using the substitution rules~\eqref{eq:metrictetrad}, we find that the relevant tetrad components can be written in terms of the PPN parameters and potentials as
\begin{subequations}\label{eq:standardppnt}
\begin{align}
\gaor{s}{P}{2}_{00} &= \gauge{U}{P}\,,\\
\gaor{s}{P}{2}_{ij} &= \gamma\gauge{U}{P}\delta_{ij}\,,\\
\gaor{s}{P}{3}_{0i} &= -\frac{1}{4}(3 + 4\gamma + \alpha_1 - \alpha_2 + \zeta_1 - 2\xi)\gauge{V}{P}_i - \frac{1}{4}(1 + \alpha_2 - \zeta_1 + 2\xi)\gauge{W}{P}_i\,,\\
\gaor{s}{P}{4}_{00} &= \frac{1}{2}(1 - 2\beta)\gauge{U}{P}^2 + \frac{1}{2}(2 + 2\gamma + \alpha_3 + \zeta_1 - 2\xi)\gauge{\Phi}{P}_1 + (1 + 3\gamma - 2\beta + \zeta_2 + \xi)\gauge{\Phi}{P}_2\nonumber\\
&\phantom{=}+ (1 + \zeta_3)\gauge{\Phi}{P}_3 + (3\gamma + 3\zeta_4 - 2\xi)\gauge{\Phi}{P}_4 - \xi\gauge{\Phi}{P}_W - \frac{1}{2}(\zeta_1 - 2\xi)\gauge{\mathfrak{A}}{P}\,.
\end{align}
\end{subequations}
Observe that only the symmetric parts of the tetrad perturbations enter the calculation of the standard PPN metric~\eqref{eq:standardppn}. Hence, these are the only components we write in terms of PPN parameters and PPN potentials. Comparing these components with the tetrad~\eqref{eq:tratetrad} written in terms of gauge-invariant components and gauge defining vector fields, we find that the vector fields which transform the tetrad~\eqref{eq:gitetrad} from the distinguished gauge to the standard PPN gauge are the same vector fields~\eqref{eq:ppngvecf} which we also found in the metric case. Recall, however, our remark from section~\ref{ssec:ginvtet} that the distinguished gauge we chose to define the gauge-invariant tetrad differs from our choice made in the metric case; this difference affects only higher order vector fields \(\order{P}{k}^{\mu}\) with \(k \geq 4\), and is thus not relevant for our calculation here. We finally find that the gauge-invariant tetrad components can be expressed as
\begin{subequations}\label{eq:ginvppntet}
\begin{align}
\order{\boldsymbol{\theta}}{2}^{\star} &= \mathbf{U}\,,\\
\order{\boldsymbol{\theta}}{2}^{\bullet} &= \gamma\mathbf{U}\,,\\
\order{\boldsymbol{\theta}}{2}^{\dagger}_{ij} &= 0\,,\\
\order{\boldsymbol{\theta}}{3}^{\diamond}_i &= -\frac{1}{2}\left(1 + \gamma + \frac{\alpha_1}{4}\right)(\mathbf{V}_i + \mathbf{W}_i)\,,\\
\order{\boldsymbol{\theta}}{4}^{\star} &= \frac{1}{4}(2 - \alpha_1 + 2\alpha_2 + 2\alpha_3)\boldsymbol{\Phi}_1 + (1 + 3\gamma - 2\beta + \zeta_2 + \xi)\boldsymbol{\Phi}_2 + (1 + \zeta_3)\boldsymbol{\Phi}_3 + (3\gamma + 3\zeta_4 - 2\xi)\boldsymbol{\Phi}_4\nonumber\\
&\phantom{=}- \xi\boldsymbol{\Phi}_W + \frac{1}{2}(1 - 2\beta)\mathbf{U}^2 + \frac{1}{4}(2 + 4\gamma + \alpha_1 - 2\alpha_2)\boldsymbol{\mathfrak{A}} + \frac{1}{4}(2 + 4\gamma + \alpha_1 - 2\alpha_2 + 2\zeta_1 - 4\xi)\boldsymbol{\mathfrak{B}}\,.
\end{align}
\end{subequations}
This concludes our construction of a gauge-invariant PPN formalism in the tetrad formulation. To illustrate its use, we apply it to an example theory in the following section.

\section{Example: scalar-tensor gravity}\label{sec:example}
We now apply the gauge-invariant PPN formalism developed in the preceding sections to a simple example theory. For the latter we chose the widely discussed scalar-tensor theory of gravity with a massless scalar field, whose PPN parameters are well known, so that we can immediately check our result. We briefly display this theory in section~\ref{ssec:exaction}. We then perturbatively solve its field equations in terms of gauge-invariant potentials in the metric formalism in section~\ref{ssec:exmetric}. Finally, we also display this solution in the tetrad formalism in section~\ref{ssec:extetrad}.

\subsection{Action and field equations}\label{ssec:exaction}
In the following we discuss a class of scalar-tensor theories of gravity, whose action is given by~\cite{Nordtvedt:1970uv}
\begin{equation}\label{eq:exaction}
S = \frac{1}{2\kappa^2}\int_Md^4x\sqrt{-g}\left(\psi R - \frac{\omega(\psi)}{\psi}\partial_{\rho}\psi\partial^{\rho}\psi\right) + S_m[g_{\mu\nu}, \chi]
\end{equation}
in Brans-Dicke like parametrization in the Jordan conformal frame. Here \(S_m\) denotes the matter part of the action, where we collectively denoted by \(\chi\) the set of matter fields. The gravitational part contains a free function \(\omega\) of the scalar field \(\psi\). Each theory of this class is defined by a particular choice of this free function \(\omega\). By variation of this action with respect to the metric and the scalar field as well as subtraction of a suitable multiple of the trace of the metric field equation one obtains the field equations
\begin{subequations}\label{eq:stgfeq}
\begin{align}
\psi R_{\mu\nu} - \nabla_{\mu}\partial_{\nu}\psi - \frac{\omega}{\psi}\partial_{\mu}\psi\partial_{\nu}\psi + \frac{g_{\mu\nu}}{4\omega + 6}\frac{d\omega}{d\psi}\partial_{\rho}\psi\partial^{\rho}\psi &= \kappa^2\left(T_{\mu\nu} - \frac{\omega + 1}{2\omega + 3}g_{\mu\nu}T\right)\,,\label{eq:stgmeteq}\\
(2\omega + 3)\square\psi + \frac{d\omega}{d\psi}\partial_{\rho}\psi\partial^{\rho}\psi &= \kappa^2T\,,\label{eq:stgscaleq}
\end{align}
\end{subequations}
where \(\square = g^{\mu\nu}\nabla_{\mu}\nabla_{\nu}\) is the d'Alembert operator on the physical spacetime \(M\). In order to perturbatively solve these equations, we also need to provide a perturbative expansion of the scalar field \(\psi\), as well as an expansion in terms of gauge-invariant quantities. First note that the relevant perturbation orders are given by
\begin{equation}
\gaor{\psi}{X}{0} = \Psi\,, \quad \gaor{\psi}{X}{2}\,, \quad \gaor{\psi}{X}{4}
\end{equation}
in a gauge \(\mathcal{X}\), where \(\Psi\) denotes the constant cosmological background value of \(\psi\). Using the gauge transformation
\begin{equation}
\gaor{\psi}{Y}{2} = \gaor{\psi}{X}{2}\,, \quad
\gaor{\psi}{Y}{4} = \gaor{\psi}{X}{4} + \order{\xi}{2}_i\gaor{\psi}{X}{2}_{,i}
\end{equation}
to a different gauge \(\mathcal{Y}\), we can define the gauge-invariant scalar field perturbations \(\order{\boldsymbol{\psi}}{k}\) as the perturbations in the distinguished gauge determined by the choice of the metric (or tetrad), so that in an arbitrary gauge we have
\begin{equation}
\gaor{\psi}{X}{2} = \order{\boldsymbol{\psi}}{2}\,, \quad
\gaor{\psi}{X}{4} = \order{\boldsymbol{\psi}}{4} + (\partial_i\order{X}{2}^{\bdiamond} + \order{X}{2}^{\diamond}_i)\order{\boldsymbol{\psi}}{2}_{,i}\,.
\end{equation}
Finally, also the function \(\omega(\psi)\) must be expanded into a Taylor series around the cosmological background. Here we introduce the shorthand notation
\begin{equation}
\omega_0 = \omega(\Psi)\,, \quad
\omega_1 = \omega'(\Psi)
\end{equation}
for the Taylor coefficients, which we assume to be of zeroth velocity order \(\mathcal{O}(0)\).

\subsection{Solution in metric formulation}\label{ssec:exmetric}
In order to illustrate the gauge-invariant metric PPN formalism detailed in section~\ref{sec:gippn}, we now apply it to the field equations~\eqref{eq:stgfeq}, pulled back to the reference spacetime \(M_0\) by an arbitrary gauge \(\mathcal{X}\). First, observe that the zeroth order \(\gaor{\psi}{X}{0} = \Psi, \gaor{g}{X}{0}_{\mu\nu} = \eta_{\mu\nu}\) indeed solves the zeroth order (vacuum) field equations. We continue with the time component of the second-order metric field equations~\eqref{eq:stgmeteq}. The corresponding equation reads
\begin{equation}\label{eq:stg2ord00}
-\frac{1}{2}\Psi\triangle\gaor{g}{X}{2}_{00} = \kappa^2\left[\gaor{T}{X}{2}_{00} + \frac{\omega_0 + 1}{2\omega_0 + 3}(\gaor{T}{X}{2}_{ii} - \gaor{T}{X}{2}_{00})\right]\,.
\end{equation}
We then express both sides of the equation by gauge-invariant quantities. On the left hand side we thus substitute the metric component \(\gaor{g}{X}{2}_{00} = \order{\mathbf{g}}{2}^{\star}\) using the relation~\eqref{eq:trametric}, while on the right hand side we have the energy-momentum tensor~\eqref{eq:traenmom} and thus
\begin{equation}
\gaor{T}{X}{2}_{00} = \order{\mathbf{T}}{2}_{00} = \order{\mathbf{T}}{2}^{\star} = \boldsymbol{\rho}\,, \quad
\gaor{T}{X}{2}_{ii} = \order{\mathbf{T}}{2}_{ii} = 3\order{\mathbf{T}}{2}^{\bullet} = 0\,.
\end{equation}
The resulting gauge-invariant equation and its solution are thus given by
\begin{equation}\label{eq:stg2sol1}
-\frac{1}{2}\Psi\triangle\order{\mathbf{g}}{2}^{\star} = \kappa^2\frac{\omega_0 + 2}{2\omega_0 + 3}\boldsymbol{\rho} \quad \Rightarrow \quad
\order{\mathbf{g}}{2}^{\star} = \frac{\kappa^2}{2\pi\Psi}\frac{\omega_0 + 2}{2\omega_0 + 3}\mathbf{U}\,.
\end{equation}
At this point it is helpful to recall the standard normalization~\eqref{eq:ginewtlim} of the gravitational constant, through which the Newtonian limit takes the form \(\order{\mathbf{g}}{2}^{\star} = 2\mathbf{U}\). We perform this normalization here by setting
\begin{equation}
\kappa^2 = 4\pi\Psi\frac{2\omega_0 + 3}{\omega_0 + 2}\,,
\end{equation}
which we will use during the remainder of this section. In the next step we consider the scalar field equation~\eqref{eq:stgscaleq} at the second velocity order, which in an arbitrary gauge \(\mathcal{X}\) reads
\begin{equation}
(2\omega_0 + 3)\triangle\gaor{\psi}{X}{2} = \kappa^2(\gaor{T}{X}{2}_{ii} - \gaor{T}{X}{2}_{00})\,.
\end{equation}
Performing the same substitution for the energy-momentum tensor as in the metric field equation above, as well as substituting \(\gaor{\psi}{X}{2} = \order{\boldsymbol{\psi}}{2}\), we then find
\begin{equation}
(2\omega_0 + 3)\triangle\order{\boldsymbol{\psi}}{2} = -\kappa^2\boldsymbol{\rho} \quad \Rightarrow \quad
\order{\boldsymbol{\psi}}{2} = \frac{\kappa^2}{4\pi(2\omega_0 + 3)}\mathbf{U} = \frac{\Psi}{\omega_0 + 2}\mathbf{U}\,.
\end{equation}
We then continue with the second-order spatial components of the field equation~\eqref{eq:stgmeteq}. In an arbitrary gauge \(\mathcal{X}\) we have
\begin{equation}
-\frac{1}{2}\Psi\left(\triangle\gaor{g}{X}{2}_{ij} - \gaor{g}{X}{2}_{00,ij} + \gaor{g}{X}{2}_{kk,ij} - \gaor{g}{X}{2}_{ik,jk} - \gaor{g}{X}{2}_{jk,ik}\right) - \gaor{\psi}{X}{2}_{,ij} = \kappa^2\left[\gaor{T}{X}{2}_{ij} - \frac{\omega_0 + 1}{2\omega_0 + 3}\delta_{ij}(\gaor{T}{X}{2}_{ii} - \gaor{T}{X}{2}_{00})\right]\,.
\end{equation}
On the right hand side we substitute the energy-momentum tensor as in the previous equations. On the left hand side we substitute the metric components using the relations~\eqref{eq:trametric}. By the virtue of the gauge-invariant decomposition we find that all occurrences of the gauge defining vector field \(X\) cancel and the resulting equation takes the simple form
\begin{equation}\label{eq:stg2ordij}
-\frac{1}{2}\Psi(\delta_{ij}\triangle\order{\mathbf{g}}{2}^{\bullet} + \order{\mathbf{g}}{2}^{\bullet}_{,ij} - \order{\mathbf{g}}{2}^{\star}_{,ij} + \triangle\order{\mathbf{g}}{2}^{\dagger}_{ij}) - \order{\boldsymbol{\psi}}{2}_{,ij} = \kappa^2\frac{\omega_0 + 1}{2\omega_0 + 3}\delta_{ij}\boldsymbol{\rho}\,.
\end{equation}
This equation may now be decomposed into its trace, a trace-free second derivative and a transverse, trace-free part. To isolate the former, we take the trace
\begin{equation}
-\frac{1}{2}\Psi(4\triangle\order{\mathbf{g}}{2}^{\bullet} - \triangle\order{\mathbf{g}}{2}^{\star}) - \triangle\order{\boldsymbol{\psi}}{2} = 3\kappa^2\frac{\omega_0 + 1}{2\omega_0 + 3}\boldsymbol{\rho}\,.
\end{equation}
To solve for \(\order{\mathbf{g}}{2}^{\bullet}\), we substitute the previously found solutions for \(\order{\mathbf{g}}{2}^{\star}\) and \(\order{\boldsymbol{\psi}}{2}\) and obtain
\begin{equation}\label{eq:stg2sol2}
-2\Psi\triangle\order{\mathbf{g}}{2}^{\bullet} = 4\kappa^2\frac{\omega_0 + 1}{2\omega_0 + 3}\boldsymbol{\rho} \quad \Rightarrow \quad
\order{\mathbf{g}}{2}^{\bullet} = \frac{\kappa^2}{2\pi\Psi}\frac{\omega_0 + 1}{2\omega_0 + 3}\mathbf{U} = 2\frac{\omega_0 + 1}{\omega_0 + 2}\mathbf{U}\,.
\end{equation}
One now easily checks that this solution also solves the trace-free second derivative part of the second-order equation~\eqref{eq:stg2ordij}, which reads
\begin{equation}
-\triangle_{ij}\left[\frac{1}{2}\Psi(\order{\mathbf{g}}{2}^{\bullet} - \order{\mathbf{g}}{2}^{\star}) + \order{\boldsymbol{\psi}}{2}\right] = 0\,.
\end{equation}
This is a direct consequence of the gauge invariance of the theory, which implies that the second-order scalar parts of the field equations are linearly dependent. We are left with the transverse, trace-free part, from which we find
\begin{equation}\label{eq:stg2sol3}
\triangle\order{\mathbf{g}}{2}^{\dagger}_{ij} = 0 \quad \Rightarrow \quad
\order{\mathbf{g}}{2}^{\dagger}_{ij} = 0\,.
\end{equation}
Next, we consider the third-order mixed components of the field equations~\eqref{eq:stgmeteq}, which read
\begin{equation}
-\frac{1}{2}\Psi(\triangle\gaor{g}{X}{3}_{0i} - \gaor{g}{X}{3}_{0j,ij} + \gaor{g}{X}{2}_{jj,0i} - \gaor{g}{X}{2}_{ij,0j}) - \gaor{\psi}{X}{2}_{,0i} = \kappa^2\gaor{T}{X}{3}_{0i}\,.
\end{equation}
Again it is the virtue of the gauge-invariant decomposition that, once we substitute the metric components~\eqref{eq:trametric}, the left hand side of these equations greatly simplifies. On the right hand side we substitute the energy-momentum tensor~\eqref{eq:traenmom} and thus
\begin{equation}
\gaor{T}{X}{3}_{0i} = \order{\mathbf{T}}{3}_{0i} = \order{\mathbf{T}}{3}^{\diamond}_i + \partial_i\order{\mathbf{T}}{3}^{\bdiamond} = -\boldsymbol{\rho}\mathbf{v}_i\,,
\end{equation}
from which we obtain
\begin{equation}
-\frac{1}{2}\Psi(\triangle\order{\mathbf{g}}{3}^{\diamond}_i + 2\order{\mathbf{g}}{2}^{\bullet}_{,0i}) - \order{\boldsymbol{\psi}}{2}_{,0i} = \kappa^2(\order{\mathbf{T}}{3}^{\diamond}_i + \partial_i\order{\mathbf{T}}{3}^{\bdiamond}) = -\kappa^2\boldsymbol{\rho}\mathbf{v}_i\,.
\end{equation}
This equation evidently splits into a pure divergence and a divergence-free part. Starting with the former, which reads
\begin{equation}
-\Psi\order{\mathbf{g}}{2}^{\bullet}_{,0i} - \order{\boldsymbol{\psi}}{2}_{,0i} = \kappa^2\partial_i\order{\mathbf{T}}{3}^{\bdiamond} = -\frac{\kappa^2}{4\pi}\mathbf{U}_{,0i}\,,
\end{equation}
we find that it is already solved identically by the second-order gauge-invariant components we determined above. This is another consequence of the gauge invariance of the theory, by which this equation becomes linearly dependent on the previously solved equations. We are thus left with the divergence-free part, which leads to
\begin{equation}\label{eq:stg3sol}
-\frac{1}{2}\Psi\triangle\order{\mathbf{g}}{3}^{\diamond}_i = \kappa^2\order{\mathbf{T}}{3}^{\diamond}_i = \frac{\kappa^2}{8\pi}\triangle(\mathbf{V}_i + \mathbf{W}_i) \quad \Rightarrow \quad
\order{\mathbf{g}}{3}^{\diamond}_i = -\frac{\kappa^2}{4\pi\Psi}(\mathbf{V}_i + \mathbf{W}_i) = -\frac{2\omega_0 + 3}{\omega_0 + 2}(\mathbf{V}_i + \mathbf{W}_i)\,.
\end{equation}
Finally, we solve the fourth-order temporal part of the metric field equation~\eqref{eq:stgmeteq}, which takes the form
\begin{multline}\label{eq:stg4ordX}
-\frac{1}{2}\gaor{\psi}{X}{2}\triangle\gaor{g}{X}{2}_{00} - \frac{1}{2}\Psi\left[\triangle\gaor{g}{X}{4}_{00} + \gaor{g}{X}{2}_{ii,00} - 2\gaor{g}{X}{3}_{0i,0i} + \frac{1}{2}\gaor{g}{X}{2}_{00,i}\left(\gaor{g}{X}{2}_{00,i} - 2 \gaor{g}{X}{2}_{ij,j} + \gaor{g}{X}{2}_{jj,i}\right) - \gaor{g}{X}{2}_{ij}\gaor{g}{X}{2}_{00,ij}\right]\\
- \gaor{\psi}{X}{2}_{,00} - \frac{1}{2}\gaor{g}{X}{2}_{00,i}\gaor{\psi}{X}{2}_{,i} - \frac{\omega_1}{4\omega_0 + 6}\gaor{\psi}{X}{2}_{,i}\gaor{\psi}{X}{2}_{,i} = \kappa^2\bigg[\gaor{T}{X}{4}_{00} - \frac{\omega_0 + 1}{2\omega_0 + 3}\gaor{g}{X}{2}_{00}\left(\gaor{T}{X}{2}_{ii} - \gaor{T}{X}{2}_{00}\right)\\
+ \frac{\omega_1}{(2\omega_0 + 3)^2}\gaor{\psi}{X}{2}\left(\gaor{T}{X}{2}_{ii} - \gaor{T}{X}{2}_{00}\right) + \frac{\omega_0 + 1}{2\omega_0 + 3}\left(\gaor{T}{X}{4}_{ii} - \gaor{T}{X}{4}_{00} - \gaor{g}{X}{2}_{ij}\gaor{T}{X}{2}_{ij} - \gaor{g}{X}{2}_{00}\gaor{T}{X}{2}_{00}\right)\bigg]\,.
\end{multline}
Again we make use of the substitution~\eqref{eq:trametric} for the metric components and~\eqref{eq:traenmom} for the energy-momentum tensor in an arbitrary gauge. After applying these substitutions the field equation~\eqref{eq:stg4ordX} takes the simpler form
\begin{multline}\label{eq:stg4ord}
-\frac{1}{2}\order{\boldsymbol{\psi}}{2}\triangle\order{\mathbf{g}}{2}^{\star} - \frac{1}{2}\Psi\left[\triangle\order{\mathbf{g}}{4}^{\star} + (\order{X}{2}^{\bdiamond}_{,i} + \order{X}{2}^{\diamond}_i)\triangle\order{\mathbf{g}}{2}^{\star}_{,i} + 3\order{\mathbf{g}}{2}^{\bullet}_{,00} + \frac{1}{2}\order{\mathbf{g}}{2}^{\star}_{,i}(\order{\mathbf{g}}{2}^{\star}_{,i} + \order{\mathbf{g}}{2}^{\bullet}_{,i}) - \order{\mathbf{g}}{2}^{\star}_{,ij}(\order{\mathbf{g}}{2}^{\bullet}\delta_{ij} + \order{\mathbf{g}}{2}^{\dagger}_{ij})\right] - \order{\boldsymbol{\psi}}{2}_{,00} - \frac{1}{2}\order{\mathbf{g}}{2}^{\star}_{,i}\order{\boldsymbol{\psi}}{2}_{,i} - \frac{\omega_1}{4\omega_0 + 6}\order{\boldsymbol{\psi}}{2}_{,i}\order{\boldsymbol{\psi}}{2}_{,i}\\
= \kappa^2\left\{\order{\mathbf{T}}{4}^{\star} + (\order{X}{2}^{\bdiamond}_{,i} + \order{X}{2}^{\diamond}_i)\order{\mathbf{T}}{2}^{\star}_{,i} + \frac{\omega_0 + 1}{2\omega_0 + 3}\left[3\order{\mathbf{T}}{4}^{\bullet} - \order{\mathbf{T}}{4}^{\star} - (\order{X}{2}^{\bdiamond}_{,i} + \order{X}{2}^{\diamond}_i)\order{\mathbf{T}}{2}^{\star}_{,i}\right] - \frac{\omega_1}{(2\omega_0 + 3)^2}\order{\boldsymbol{\psi}}{2}\order{\mathbf{T}}{2}^{\star}\right\}
\end{multline}
in terms of the gauge-invariant quantities. In this case we find that also the components \(\order{X}{2}^{\bdiamond}\) and \(\order{X}{2}^{\diamond}_i\) of the gauge defining vector fields appear on both sides of the field equations. This is not surprising, since the field equations~\eqref{eq:stg4ordX} are expressed in an arbitrary gauge \(\mathcal{X}\), and thus differ from the gauge-invariant field equations by a term of the form \((\order{X}{2}^{\bdiamond}_{,i} + \order{X}{2}^{\diamond}_i)\mathbf{E}_{,i}\), where \(\mathbf{E}\) are the second-order field equations~\eqref{eq:stg2ord00}. Indeed, we find that the occurrence of \(X\) in the field equation~\eqref{eq:stg4ord} is exactly of this form. Imposing that the second-order field equation is already satisfied by the solution we constructed above, we may thus drop these terms and retain the gauge-invariant fourth-order field equation
\begin{multline}
-\frac{1}{2}\order{\boldsymbol{\psi}}{2}\triangle\order{\mathbf{g}}{2}^{\star} - \frac{1}{2}\Psi\left[\triangle\order{\mathbf{g}}{4}^{\star} + 3\order{\mathbf{g}}{2}^{\bullet}_{,00} + \frac{1}{2}\order{\mathbf{g}}{2}^{\star}_{,i}(\order{\mathbf{g}}{2}^{\star}_{,i} + \order{\mathbf{g}}{2}^{\bullet}_{,i}) - \order{\mathbf{g}}{2}^{\star}_{,ij}(\order{\mathbf{g}}{2}^{\bullet}\delta_{ij} + \order{\mathbf{g}}{2}^{\dagger}_{ij})\right] - \order{\boldsymbol{\psi}}{2}_{,00} - \frac{1}{2}\order{\mathbf{g}}{2}^{\star}_{,i}\order{\boldsymbol{\psi}}{2}_{,i} - \frac{\omega_1}{4\omega_0 + 6}\order{\boldsymbol{\psi}}{2}_{,i}\order{\boldsymbol{\psi}}{2}_{,i}\\
= \kappa^2\left[\order{\mathbf{T}}{4}^{\star} + \frac{\omega_0 + 1}{2\omega_0 + 3}\left(3\order{\mathbf{T}}{4}^{\bullet} - \order{\mathbf{T}}{4}^{\star}\right) - \frac{\omega_1}{(2\omega_0 + 3)^2}\order{\boldsymbol{\psi}}{2}\order{\mathbf{T}}{2}^{\star}\right]\,.
\end{multline}
In order to solve this equation for the final remaining metric component \(\order{\mathbf{g}}{4}^{\star}\), we move all other terms to the right hand side and insert the previously found lower order solutions, as well as the energy-momentum tensor~\eqref{eq:enmomppnpot}. This yields the equation
\begin{equation}
\begin{split}
\triangle\order{\mathbf{g}}{4}^{\star} &= 8\pi\left(\frac{3}{\omega_0 + 2} + \frac{\omega_1\Psi}{(2\omega_0 + 3)(\omega_0 + 2)^2}\right)\boldsymbol{\rho}\mathbf{U} - 8\pi\frac{2\omega_0 + 3}{\omega_0 + 2}\boldsymbol{\rho}\mathbf{v}^2 - 8\pi\boldsymbol{\rho}\boldsymbol{\Pi} - 24\pi\frac{\omega_0 + 1}{\omega_0 + 2}\mathbf{p} - 2\frac{3\omega_0 + 4}{\omega_0 + 2}\mathbf{U}_{,00}\\
&\phantom{=}- \left(4 + \frac{\omega_1\Psi}{(2\omega_0 + 3)(\omega_0 + 2)^2}\right)\mathbf{U}_{,i}\mathbf{U}_{,i}\\
&= \frac{3\omega_0 + 4}{\omega_0 + 2}\triangle(\boldsymbol{\mathfrak{A}} + \boldsymbol{\mathfrak{B}}) + \triangle\boldsymbol{\Phi}_1 + \left(\frac{4\omega_0 + 2}{\omega_0 + 2} - \frac{\omega_1\Psi}{(2\omega_0 + 3)(\omega_0 + 2)^2}\right)\triangle\boldsymbol{\Phi}_2 + 3\triangle\boldsymbol{\Phi}_3 + 6\frac{\omega_0 + 1}{\omega_0 + 2}\triangle\boldsymbol{\Phi}_4\\
&\phantom{=}- 2\left(1 + \frac{\omega_1\Psi}{4(2\omega_0 + 3)(\omega_0 + 2)^2}\right)\triangle\mathbf{U}^2\,,
\end{split}
\end{equation}
together with the straightforward solution
\begin{equation}\label{eq:stg4sol}
\begin{split}
\order{\mathbf{g}}{4}^{\star} &= \frac{3\omega_0 + 4}{\omega_0 + 2}(\boldsymbol{\mathfrak{A}} + \boldsymbol{\mathfrak{B}}) + \boldsymbol{\Phi}_1 + \left(\frac{4\omega_0 + 2}{\omega_0 + 2} - \frac{\omega_1\Psi}{(2\omega_0 + 3)(\omega_0 + 2)^2}\right)\boldsymbol{\Phi}_2 + 3\boldsymbol{\Phi}_3 + 6\frac{\omega_0 + 1}{\omega_0 + 2}\boldsymbol{\Phi}_4\\
&\phantom{=}- 2\left(1 + \frac{\omega_1\Psi}{4(2\omega_0 + 3)(\omega_0 + 2)^2}\right)\mathbf{U}^2\,.
\end{split}
\end{equation}
By comparison of the full solution~\eqref{eq:stg2sol1}, \eqref{eq:stg2sol2}, \eqref{eq:stg2sol3}, \eqref{eq:stg3sol} and~\eqref{eq:stg4sol} with the gauge-invariant PPN metric~\eqref{eq:ginvppnmet} one now finds that it does indeed possess the standard PPN form, where the PPN parameters are given by
\begin{equation}\label{eq:exppnpar}
\gamma = \frac{\omega_0 + 1}{\omega_0 + 2}\,, \quad
\beta = 1 + \frac{\omega_1\Psi}{4(2\omega_0 + 3)(\omega_0 + 2)^2}\,, \quad
\alpha_1 = \alpha_2 = \alpha_3 = \zeta_1 = \zeta_2 = \zeta_3 = \zeta_4 = \xi = 0\,.
\end{equation}
This is of course the well-known post-Newtonian limit of scalar-tensor gravity with a massless scalar field~\cite{Nordtvedt:1970uv}.

We finally remark that instead of using the full metric perturbations~\eqref{eq:trametric} in an arbitrary gauge \(\mathcal{X}\) and the corresponding energy-momentum tensor~\eqref{eq:traenmom} we could also have assumed \(\order{X}{k} = 0\) from the beginning, thus effectively working in the distinguished gauge \(\mathcal{X} = \mathcal{S}\) used to define the gauge-invariant metric components, since the gauge defining vector fields \(\order{X}{k}\) do not contribute to the field equations, and so any gauge choice is valid. However, we chose to work in an arbitrary gauge here in order to demonstrate this fact, i.e., to explicitly show that the vector fields \(\order{X}{k}\) cancel and the field equations yield gauge-invariant quantities only.

\subsection{Solution in tetrad formulation}\label{ssec:extetrad}
Instead of solving the PPN expansion of the field equations~\eqref{eq:stgfeq} for the metric perturbations~\eqref{eq:gimetric}, one may also make use of the relations~\eqref{eq:metrictetrad} and express the field equations through tetrads instead, and then solve for the tetrad components~\eqref{eq:gitetrad} in order to determine the PPN parameters from the expression~\eqref{eq:ginvppntet}. We will not perform this procedure here in detail, and only note that the result is given by
\begin{gather}
\order{\boldsymbol{\theta}}{2}^{\star} = \mathbf{U}\,, \quad
\order{\boldsymbol{\theta}}{2}^{\bullet} = \frac{\omega_0 + 1}{\omega_0 + 2}\mathbf{U}\,, \quad
\order{\boldsymbol{\theta}}{2}^{\dagger}_{ij} = 0\,, \quad
\order{\boldsymbol{\theta}}{3}^{\diamond}_i = -\frac{2\omega_0 + 3}{2\omega_0 + 4}(\mathbf{V}_i + \mathbf{W}_i)\,, \quad
\order{\boldsymbol{\theta}}{4}^{\star} = \frac{3\omega_0 + 4}{2\omega_0 + 4}(\boldsymbol{\mathfrak{A}} + \boldsymbol{\mathfrak{B}})\nonumber\\
+ \frac{1}{2}\boldsymbol{\Phi}_1 + \left(\frac{2\omega_0 + 1}{\omega_0 + 2} - \frac{\omega_1\Psi}{2(2\omega_0 + 3)(\omega_0 + 2)^2}\right)\boldsymbol{\Phi}_2 + \frac{3}{2}\boldsymbol{\Phi}_3 + 3\frac{\omega_0 + 1}{\omega_0 + 2}\boldsymbol{\Phi}_4 - \left(\frac{1}{2} + \frac{\omega_1\Psi}{4(2\omega_0 + 3)(\omega_0 + 2)^2}\right)\mathbf{U}^2\,.
\end{gather}
By comparison with the gauge-invariant PPN tetrad~\eqref{eq:ginvppntet} one obtains the same PPN parameters~\eqref{eq:exppnpar} as by using the metric formulation detailed above. Of course, in the case of the scalar-tensor theory detailed in section~\ref{ssec:exaction} these two approaches are exactly equivalent, since the field equations~\eqref{eq:stgfeq} are fully expressed in terms of the metric, and so there is no benefit in using the tetrad formulation. Here we use it only as a proof of concept. Its full virtue can be exploited by applying the formalism to theories which have a more natural formulation in terms of tetrads, such as bimetric gravity~\cite{Hinterbichler:2012cn,Schmidt-May:2015vnx} or teleparallel gravity~\cite{Einstein:1928,Aldrovandi:2013wha}, where also the remaining, antisymmetric components of the tetrad perturbations will enter the field equations as auxiliary fields and must be solved for~\cite{Ualikhanova:2019ygl,Emtsova:2019qsl,Flathmann:2019khc}. We will not discuss the details of this procedure here, as this would exceed the scope of this article.

\section{Conclusion}\label{sec:conclusion}
We have applied the theory of gauge-invariant higher order perturbations to the PPN formalism and developed a formulation which is independent of the choice of the gauge, i.e., the coordinate system in which the post-Newtonian approximation is performed. We provided explicit formulas for the metric perturbations and energy-momentum tensor in an arbitrary gauge and expressed their gauge-invariant components in terms of the well-known PPN parameters and PPN potentials. In addition to the standard metric formulation, we also devised a tetrad formulation, which is more suitable for gravity theories in which the fundamental field is a tetrad. We finally demonstrated the practical use of the gauge-invariant PPN formalism by applying it to an example class of scalar-tensor gravity theories and re-deriving its PPN parameters.

Possibilities for future research arise mainly from applying our formalism to gravity theories and deriving their PPN parameters. The virtue of the gauge-invariant approach lies in the fact that it isolates the physical, gauge-invariant degrees of freedom, while removing any gauge or coordinate dependence from the equations, which may otherwise clutter the calculation. This potentially leads to a significant simplification of the equations to be solved. Further, it removes the arbitrariness in a priori choosing a gauge, which is necessary to solve the field equations in the standard PPN formalism. This fact may be used by employing computer algebra in order to automatize solving the field equations, without any further input such as the choice of gauge.

Various extensions and modifications of this formalism are possible. For example, one may adapt the gauge-invariant formalism to a modified version of the PPN formalism, which makes use of a different density variable~\cite[Sec. 4]{Will:2018bme}. Another possibility is to consider theories with more than one dynamical metric (or tetrad) and perform a post-Newtonian expansion for both metrics~\cite{Clifton:2010hz,Hohmann:2010ni,Hohmann:2013oca}. In this case one must pay attention to the fact that once the gauge-invariant variables for one of the metrics are chosen, following the prescription shown in this article, there is no further possibility to eliminate certain components of other dynamical metrics by gauge transformations, and so all components must be expressed in terms of gauge-invariant quantities. Further possibilities are including the Vainshtein mechanism~\cite{Avilez-Lopez:2015dja} or the time dependence of PPN parameters in a cosmological background spacetime~\cite{Sanghai:2016tbi}. Finally, additional PPN parameters and potentials may be included, in order to accommodate, e.g., for massive fields leading to Yukawa-type terms~\cite{Zaglauer:1990yh,Helbig:1991pk}, terms of higher derivative order~\cite{Gladchenko:1990nw} or parity-violating terms as in Chern-Simons gravity~\cite[Sec. 5.6]{Will:2018bme}.

Further, one may include even higher perturbation orders, and apply the gauge-invariant perturbation theory to the calculation of gravitational radiation from compact sources~\cite{Blanchet:2013haa}. While the conventional approach heavily relies on the choice of the gauge, also here a gauge-invariant formulation may lead to new insights and a simplified approach.

\begin{acknowledgments}
The author gratefully acknowledges the full support by the Estonian Research Council through the Personal Research Funding project PRG356, as well as the European Regional Development Fund through the Center of Excellence TK133 ``The Dark Side of the Universe''.
\end{acknowledgments}

\bibliography{gippn}
\end{document}